\documentstyle[11pt]{report}
 \newcommand{\be}{\begin{equation}}
\newcommand{\ee}{\end{equation}}
\newcommand{\bea}{\begin{eqnarray}}
\newcommand{\eea}{\end{eqnarray}}
\newcommand{\lb}{\label}
\newcommand{\da}{\dagger}
\begin{document}
\begin{titlepage}
\begin{flushright}
 ZU-TH 8/93\\
ETH-TH/93-17\\
hep-th/9306161
\end{flushright}
\begin{center}
\vfill
{\large\bf  FUNCTIONAL SCHR\"ODINGER EQUATION FOR FERMIONS IN EXTERNAL
   GAUGE FIELDS}
\vskip 0.7cm
{\bf Claus Kiefer$^*$}
\vskip 0.4cm
Institute for Theoretical Physics, University of Z\"urich,
Sch\"onberggasse 9,\\CH-8001 Z\"urich, Switzerland
\vskip 0.7cm
{\bf Andreas Wipf}
\vskip 0.4cm
Institute for Theoretical Physics, ETH - H\"onggerberg,\\
CH-8093 Z\"urich, Switzerland
\end{center}
\vfill
\begin{center}
{\bf Abstract}
\end{center}
\begin{quote}
We discuss the functional Schr\"odinger picture for fermionic fields in
external gauge fields for both stationary and time - dependent problems.  We
give formal results for the ground state and the solution of the time -
dependent Schr\"odinger equation for QED in arbitrary dimensions, while more
explicit results are obtained in two dimensions. For both the massless and
massive Schwinger model we give an explicit expression for the ground state
functional as well as for the expectation values of energy, electric and axial
charge. We also give the corresponding results for non-abelian fields. We solve
the functional Schr\"odinger equation for a constant external field in four
dimensions and obtain the amount of particle creation. We solve the
Schr\"odinger equation for arbitrary external fields for massless QED in two
dimensions and make a careful discussion of the anomalous particle creation
rate. Finally, we discuss some subtleties connected with the interpretation of
the quantized Gauss constraint.

 \end{quote}
\vfill
\begin{center}
{\em submitted to Annals of Physics}
 \end{center}
\vfill
$^*$ Present address: Fakult\"at f\"ur Physik, Universit\"at Freiburg,
  Hermann-Herder-Str. 3, D-79104 Freiburg, Germany.

\end{titlepage}

\chapter{Introduction}

Quantum field theory has been successfully applied in various formal schemes.
High energy experiments often deal with scattering processes, where the
interaction can be treated as a perturbation. It is
therefore not surprising that their theoretical description
makes use
of methods, such as the S-matrix or path integral approach, which are well
suited for perturbation theory. There is, however, an increasing area of
applications where such methods are of limited use. Among these are the
confinement problem in QCD, anomalies, and large gauge transformations, to
mention
only a few. In these applications the language of wave functionals and the
functional Schr\"odinger equation has provided valuable insights (see, e. g.,
\cite{Ja} and \cite{Yee} for a review). One big advantage of this language is
that the intuitive picture of evolving wave functions, so successful in quantum
mechanics, can be extended to problems in field theory. It is of course still
an open problem whether the existence of the Schr\"odinger picture can
be proven rigorously. At least in the case of
 renormalizable scalar field theories
it has been demonstrated that a functional Schr\"odinger equation
with respect to a global time parameter exists
at each order of perturbation theory \cite{Sy}. For arbitrary local
time variations an explicit calculation has verified the validity
of the Schr\"odinger equation up to two loops \cite{Av}.

An important field of application is quantum gravity. Since quantum general
relativity is nonrenormalizable at the perturbative level, one has to develop
nonperturbative methods, provided the theory is viable at all. There have been
remarkable developments in canonical quantum gravity in recent years which have
so far culminated in the discovery, by using the functional Schr\"odinger
picture, of  exact
formal solutions to all constraint equations  \cite{As}. The use of
wave functionals has also been useful in performing semiclassical
approximations, for example in the derivation of formal correction terms to the
 Schr\"odinger equation from quantum gravity \cite{KS}.  It may thus
turn out to be very useful for later applications to explore the
 potentialities
of the functional Schr\"odinger picture in ordinary field theory.

In a recent paper \cite{Ki} one of us has
 discussed various aspects of scalar QED
in this framework,
such as the semiclassical approximation and external field problems like
particle creation.
 In the present paper we extend this work to the coupling of
fermionic matter to gauge fields. Apart from the last section we limit
ourselves to the case where the gauge field can be treated semiclassically,
i.e. we discuss the functional Schr\"odinger equation for the fermionic wave
functional in a prescribed external gauge field. Most of our work deals with
QED but we also give some results for the non-abelian case.

We start by giving a brief review of the functional Schr\"odinger
equation for fermions
following, with elaborations, the work of Floreanini and Jackiw \cite{FJ}
(sections~2.1 and 2.2). Gaussian states are used as generalized vacuum states,
but contrary to the bosonic case one has to fix a filling prescription for the
Dirac sea to select a particular vacuum. Section~2.3 is concerned with the
time-dependent Schr\"odinger equation. We give its formal solution for
arbitrary external fields in terms of solutions of the (first-quantized) Dirac
equation.

We then proceed to calculate the exact ground state for arbitrary external
fields in two dimensional QED in both the massless and the massive case
(section~3). We give explicit expressions for the expectation values of the
Hamiltonian, the electric charge, and the axial charge with respect to this
ground state. Regularization is performed through gauge-invariant point
splitting. All results are given for the case of finite as well as infinite
space intervals. The finite case allows a careful discussion of the dependence
of the Casimir energy on the chosen boundary conditions.

The extension to non-abelian fields in two dimensions is straightforward
(section~4). We give the exact ground state as well as the expectations values
for the Hamiltonian, the electric and axial charges.

We then proceed to discuss applications of the time - dependent Schr\"odinger
equation (section~5). The particle creation rate for constant external electric
fields is calculated in this framework
and the classical result found by Schwinger is recovered
 (section~5.1). In the massless case in
two dimensions we calculate the anomalous particle production rate for
arbitrary external fields. Its interpretation in the functional language is
very transparent -- the anomalous production rate is basically due to the
dependence of the filling prescription on the external field (section~5.2).

In the final section
we go beyond the external field approximation and discuss briefly some
subtleties connected with the interpretation of Gauss law (section~6). We show
that, except for the case when anomalies violating gauge invariance are
present, the interpretation of the Gauss constraint as a generator of gauge
transformation can be rescued even if it does no longer annihilate gauge
invariant states. We also present a brief outlook on possible future work.

  \chapter{Functional Schr\"odinger equation for fermions}

\section{Commutation relations and inner product}

In this section we give a brief review of the canonical formalism for QED and
the functional Schr\"odinger picture. Unless otherwise stated, the dimension
$D$ of spacetime is left arbitrary. The Lagrangian density is given by
\be {\cal L}=-\frac{1}{4}F_{\mu\nu}F^{\mu\nu}+i\bar{\psi}(D_{\mu}\gamma^{\mu}
  -m)\psi,
      \lb{2.1} \ee
where
\be D_{\mu}=\partial_{\mu}+ieA_{\mu} \lb{2.2} \ee
is the covariant derivative associated with the electromagnetic potential
$A_{\mu}$. The canonical momenta read
\be \pi_0=0,\ \ \pi_i=F_{i0}\equiv E^i, \ \ \pi_{\psi}=i\psi^{\dagger} \lb{2.3}
\ee
so that the total Hamiltonian is given by
\bea H &=& \int dx\left(\frac{1}{2}{\bf E}^2 +\frac{1}{4}F^{ij}F_{ij}\right)
        +\int dxdy\psi^{\dagger}(x)h(x,y)\psi(y) \nonumber\\
     & & \ +\int dx A^0(e\psi^{\dagger}\psi-{\bf\nabla E}), \lb{2.4} \eea
where
\be h(x,y) = -i\gamma^0\gamma^i\frac{\partial}{\partial
x^i}\delta(x-y)+\gamma^0(m+
     e\gamma^iA_i)\delta(x-y)
      \lb{2.5} \ee
plays the role of a {\em first quantized} Dirac Hamiltonian in an external
electromagnetic field. We will denote with $h_{(0)}$ the first quantized
Hamiltonian without external field.
We note that $x$ and $y$ is a shorthand notation for a vector in $(D-1)$
dimensional space, and the metric convention for spacetime is
diag$(1,-1-1,...)$.
Variation of (\ref{2.4}) with respect to $A^0$ yields the Gauss constraint
\be {\bf \nabla E}=e\psi^{\dagger}\psi. \lb{2.6} \ee
In the following we use the gauge condition $A^0=0$. The commutation relations
read
\be [A_i(x),E^j(y)]=i\delta^j_i\delta(x-y) \lb{2.7} \ee
for the electromagnetic field, and
\be
\{\psi_{\alpha}(x),\psi_{\beta}^{\dagger}(y)\}=\delta_{\alpha\beta}\delta(x-y)
\lb{2.8} \ee
for the fermion fields. All other commutators (anticommutators) vanish.

In the functional Schr\"odinger picture we represent these commutation
relations by acting with the field operators on physical states
$\Psi[u,u^{\dagger},{\bf A}]$ according to
\bea E^j & \to & \frac{1}{i}\frac{\delta}{\delta A_j}, \lb{2.9} \\
       \psi_{\alpha} & \to &
\frac{1}{\sqrt{2}}\left(u_{\alpha}+\frac{\delta}{\delta
u^{\dagger}_{\alpha}}\right), \lb{2.10} \\
   \psi_{\alpha}^{\dagger} & \to &
\frac{1}{\sqrt{2}}\left(u_{\alpha}^{\dagger}+\frac{\delta}
   {\delta u_{\alpha}}\right), \lb{2.11} \eea
and ${\bf A}$ is represented by multiplication.
Note that $u_{\alpha}$ and $u_{\alpha}^{\dagger}$ are Grassmann variables, and
$\Psi$ is {\em not} an eigenstate of either $\psi$ or $\psi^{\dagger}$. An
alternative representation has been used, for example, in \cite{Du},
where $\psi$ is represented, as in the bosonic case, by multiplication
with $u$, and $\psi^{\dagger}$ is represented by $\delta/\delta u$.
Since, however, the Hermitean conjugate of $u$ in that representation
is not given by $u^{\dagger}$, but by $\delta/\delta u$, we find it
easier for our discussion to resort to the representation (2.10)
and (2.11).

The Grassmann character of the fermion fields requires a careful treatment of
the inner product \cite{FJ}. If one defines the inner product by the functional
integration (we do in the following not explicitly write out the
electromagnetic
field and the spinor indices)
\be \langle\Psi_1\vert\Psi_2\rangle\equiv\int{\cal D}u^{\dagger}{\cal
D}u\Psi_1^*\Psi_2
      =\langle\Psi_2\vert\Psi_1\rangle^*, \lb{2.12} \ee
the dual $\Psi^*$ of a state $\Psi$ is not given by ordinary complex
conjugation, but by the expression
\be \Psi^*[u,u^{\dagger}]=\int{\cal D}\bar{u}^{\dagger}{\cal
D}\bar{u}\exp\left(\bar{u}u^{\dagger}+\bar{u}^{\dagger}u\right)
\bar{\Psi}[\bar{u},\bar{u}^{\dagger}]. \lb {2.13} \ee
Here, $\bar{\Psi}$ is the hermitean conjugate of $\Psi$. We have used a compact
notation, i. e.,  $\bar{u}u\equiv\int dx\bar{u}_{\alpha}(x)u_{\alpha}(x)$, etc.
Note the analogy to the Bargmann representation for the harmonic oscillator in
quantum mechanics.

A special role is played by Gaussian states,
\be \Psi=\exp\left(u^{\dagger}\Omega u\right), \lb{2.14} \ee
since this generalizes the notion of a Fock vacuum; $\Omega$ is sometimes
called the ``covariance."

If we apply the above rules to such a state we find
\be \bar{\Psi}[\bar{u},\bar{u}^{\dagger}]=\exp\left(\bar{u}^{\dagger}
\Omega^{\dagger}\bar{u}\right),
    \lb{2.15} \ee
and for the dual, applying the familiar rules of Grassmann integration,
\bea \Psi^*[u,u^{\dagger}] & = & \int {\cal D}\bar{u}^{\dagger}{\cal D}\bar{u}
\exp\left(\bar{u}u^{\dagger}+\bar{u}^{\dagger}u+\bar{u}^{\dagger}
\Omega^{\dagger}\bar{u}\right)
        \nonumber\\
     & = &
\mbox{det}(-\Omega^{\dagger})\exp\left(u^{\dagger}
(\Omega^{\dagger})^{-1}u\right).
     \lb{2.16} \eea
One then finds for $\langle\Psi\vert\Psi\rangle$ the expression
\bea   \langle\Psi\vert\Psi\rangle & = & \mbox{det}(-\Omega^{\dagger})\int{\cal
D}u^{\dagger}{\cal
D}u\exp\left(u^{\dagger}\left[(\Omega^{\dagger})^{-1}+\Omega\right]u\right)
  \nonumber\\
 & = & \mbox{det}(1+\Omega^{\dagger}\Omega). \lb{2.17} \eea
An important difference to the bosonic case is the fact that the state
$\Psi[u,u^{\dagger}]$ is {\em not} an overlap with fields states, $
\Psi[u,u^{\dagger}]\neq\langle u,u^{\dagger}\vert\Psi\rangle$,
since the inner product is an ordinary number, whereas $\Psi$ can be expanded
in terms of Grassmann variables.

\section{Solution of the stationary Schr\"odinger equation}

In this section we look for the {\em ground state} of the Dirac Hamiltonian in
an external electromagnetic field, i. e., we solve the stationary Schr\"odinger
equation
\be \left(\int dxdy\psi^{\dagger}(x)h(x,y)\psi(y)\right)\Psi\equiv
H_{\psi}\Psi=E_0\Psi. \lb{2.18} \ee
If $\psi_n$ are the eigenmodes of the first quantized Hamiltonian $h$,
\be h\psi_n=E_n\psi_n, \lb{2.19} \ee
we can expand the field operators $\psi$ and $\psi^{\dagger}$ as
\be \psi=\sum_n a_n\psi_n, \ \ \psi^{\dagger}=\sum_n
a_n^{\dagger}\psi_n^{\dagger}, \lb{2.20} \ee
where $a_n$ ($a_n^{\dagger}$) is the usual annihilation (creation) operator.
Then,
\be H_{\psi}=\sum_n E_n a_n^{\dagger}a_n. \lb{2.21} \ee
We can also expand $u$ and $u^{\dagger}$ in terms of these eigenmodes
\be u(x)=\sum_n u_n\psi_n(x), \ \ u^{\dagger}(x)=\sum_n
u_n^{\dagger}\psi_n^{\dagger}(x). \lb{2.22} \ee
Note that
\be \frac{\delta}{\delta u(x)}=\sum_n\psi_n^{\dagger}(x)\frac{\delta}{\delta
u_n} \lb{2.23} \ee
 to guarantee that $\delta u(x)/\delta u(y)=\delta(x-y)$.
Inserting these expansions into the expression for $H_{\psi}$, we find
\be H_{\psi}=\frac{1}{2}\sum_n E_n\left(u_n^{\dagger} +\frac{\delta}{\delta
u_n}\right)\left(u_n +
    \frac{\delta}{\delta u_n^{\dagger}}\right). \lb{2.24} \ee
We want to apply this Hamiltonian on the Gaussian state (\ref{2.14}). To that
purpose we note that
\be u^{\dagger}\Omega u=\sum_{n,m}u_n^{\dagger}\Omega_{nm}u_m \lb{2.25} \ee
with
\be \Omega(x,y)=\sum_{n,m}\Omega_{nm}\psi_n(x)\psi_m^{\dagger}(y). \lb{2.26}
\ee
We then find
\bea H_{\psi}\Psi & = & \frac{1}{2}\mbox{Tr}h(1+\Omega)\Psi \nonumber\\
    & & \
+\frac{1}{2}\sum_{k,l,n}u_n^{\dagger}(\delta_{nk}-\Omega_{nk})
E_k(\delta_{kl}+\Omega_{kl})u_l
  \Psi. \lb{2.27} \eea
Upon comparison with (\ref{2.18}) we see that the ground state energy is given
by
\be E_0=\frac{1}{2}\mbox{Tr}h(1+\Omega)=\frac{1}{2}\sum_n E_n(1+\Omega_{nn}),
\lb{2.28} \ee
and that, since the second term in (\ref{2.27}) must vanish, the elements of
$\Omega_{nn}$ are given by
\be \Omega_{nm}=\pm\delta_{nm}. \lb{2.29} \ee
There still remains some arbitrariness how one distributes the numbers $1$ and
$-1$ among the elements of $\Omega$. This arbitrariness can be removed by the
use of the annihilation operators introduced above. We have
\bea a_n^{\dagger}a_n\Psi & = & \left(u_n^{\dagger} +\frac{\delta}{\delta
u_n}\right)\left(u_n +\frac{\delta}{\delta u_n^{\dagger}}\right)\Psi
\nonumber\\
 & = & \frac{1}{2}(1+\Omega_{nn})\Psi. \lb{2.30} \eea
We demand that the ground state $\Psi$ be annihilated by $a_n$ for {\em
positive} energies $E_n$, i. e., \be a_n^{\dagger}a_n\Psi =\left\{
\begin{array}{ll}
                                             0 & \mbox{if $
\Omega_{nn}=-1$}\leftrightarrow E_n >0 \\
                                              \Psi & \mbox{if $
\Omega_{nn}=+1$}\leftrightarrow E_n <0
                                            \end{array} \right. \lb{2.31} \ee
This selects a specific ground state and is equivalent to say, in a more
heuristic language, that a specific prescription for the filling of the Dirac
sea has been chosen.
{}From (\ref{2.26}) we thus find for the covariance
\be
\Omega(x,y)=\sum_{E_n<0}\psi_n(x)\psi_n^{\dagger}(y)-
\sum_{E_n>0}\psi_n(x)\psi_n^{\dagger}(y). \lb{2.32}       \ee
It is very convenient, and we will make extensive use of it later on, to
express this relation in terms of {\em projectors},
\be \Omega\equiv P_- - P_+ , \lb{2.33} \ee
where
\be P_{\pm}\equiv\frac{1\mp\Omega}{2} \lb{2.34} \ee
project on positive and negative energies, respectively:
\be P_+P_-=P_-P_+=0, \ P_+^2=P_+, \ P_-^2=P_-, \ P_+ + P_- =1. \lb{2.35} \ee
We also note the operator expression for $\Omega$, which follows from the
vanishing of the second term in (\ref{2.27}), reads:
\be \frac{1}{4}(1-\Omega)h(1+\Omega)=0=P_+ h P_-. \lb{2.36} \ee
In case that the external electromagnetic field vanishes we can give easily an
explicit expression for $\Omega$. In momentum space, the solution corresponding
to the filling prescription (\ref{2.31}) reads
\be \Omega_{(0)}(p,p')=-\frac{h_{(0)}}{\sqrt{p^2+m^2}}\delta(p-p'), \lb{2.37}
\ee
where $h_{(0)}$ is the ${\bf A}$-independent part of (\ref{2.5}).
This can most easily be seen by calculating the vacuum energy $E_0$. From
(\ref{2.28}) we have, since $h_{(0)}$ has vanishing trace,
\bea E_0 &=& \frac{1}{2}\mbox{Tr}h_{(0)}\Omega_{(0)}=\frac{1}{2}\sum_n
E_n\Omega_{nn}=
       -\frac{1}{2}\sum_n\vert E_n\vert \nonumber\\
        &=& -\frac{1}{2}\mbox{Tr}\sqrt{p^2+m^2}=-\frac{1}{2}\frac{V}{(2\pi)^3}
      \int d^3p\sqrt{p^2+m^2}. \lb{2.38} \eea
Use has been made here of the fact that the square of $h$ is given by
\be h^2_{(0)}=p^2+m^2. \lb{2.39} \ee
For later use we give the explicit result for two and four spacetime
dimensions. In two dimensions we have, in the chiral representation,
\be \Omega_{(0)}=-\frac{1}{\sqrt{p^2+m^2}}\left( \begin{array}{cc}
                         -p & m \\ m & p \end{array}\right), \lb{2.40} \ee
and in the Dirac representation
\be \Omega_{(0)}=\frac{1}{\sqrt{p^2+m^2}}\left( \begin{array}{cc}
                         -m & p \\ p & m \end{array}\right). \lb{2.41} \ee
In the four dimensional case we have, in the Dirac representation,
\be \Omega_{(0)}=-\frac{1}{\sqrt{p^2+m^2}}\left( \begin{array}{cc}
                         m & \sigma\cdot p \\ \sigma\cdot p & -m
\end{array}\right), \lb{2.42} \ee
where $\sigma$ are the Pauli matrices.

We conclude this section with a discussion of the two-point function\\
$\langle\psi_{\alpha}(x)
\psi_{\beta}^{\dagger}(y)\rangle$, where the expectation value is computed with
respect to the above ground state. For this we need the two-point function of
$uu^{\dagger}$ which we now calculate, using (\ref{2.14}) and (\ref{2.16}),
\bea & & \frac{\langle
u_{\alpha}(x)u_{\beta}^{\dagger}(y)\rangle}{\langle\Psi\vert\Psi\rangle}=
        \frac{\mbox{det}(-\Omega^{\dagger})}{\langle\Psi\vert\Psi\rangle}\int
{\cal D}u^{\dagger}{\cal D} u
         u_{\alpha}(x)u_{\beta}^{\dagger}(y)\nonumber\\
& & \ \ \ \cdot
\exp\left(u^{\dagger}[(\Omega^{\dagger})^{-1}+\Omega]u\right) \lb{2.43}\\
 & = & \
\frac{\mbox{det}(-\Omega^{\dagger})}{\langle\Psi\vert\Psi\rangle}
\frac{\delta^2}{\delta\eta_{\alpha}(x)
  \delta\eta_{\beta}^{\dagger}(y)}\int {\cal D} u^{\dagger}{\cal D}
u\nonumber\\
& & \ \ \ \cdot
          \exp\left(u^{\dagger}[(\Omega^{\dagger})^{-1}+\Omega]u+\eta
u+\eta^{\dagger}u^{\dagger}\right)
         \vert_{\eta=\eta^{\dagger}=0} \nonumber\\
 & & \
=\frac{\mbox{det}(1+\Omega^{\dagger}\Omega)}
{\langle\Psi\vert\Psi\rangle}\frac{\delta^2}
{\delta\eta_{\alpha}(x)
\delta\eta_{\beta}^{\dagger}(y)}\exp\left(\eta[(\Omega^{\dagger})^{-1}
+\Omega]^{-1}
  \eta^{\dagger}\right)\vert_{\eta=\eta^{\dagger}=0} \nonumber\\
& & \ =-[(\Omega^{\dagger})^{-1}+\Omega]^{-1}_{\alpha\beta}(x,y), \lb{2.44}
\eea
where (\ref{2.17}) has been used. In the present case, where
$\Omega=\Omega^{\dagger}$ and $\Omega^2=1$,
this reads
\be \frac{\langle
u_{\alpha}(x)u_{\beta}^{\dagger}(y)\rangle}{\langle\Psi\vert\Psi\rangle}=
       -\frac{1}{2}\Omega_{\alpha\beta}(x,y). \lb{2.45} \ee
If we apply $\psi_{\alpha}(x)
\psi_{\beta}^{\dagger}(y)$ on the ground state, we find
\begin{eqnarray*} & & \psi_{\alpha}(x)
\psi_{\beta}^{\dagger}(y)\Psi=\frac{1}{2}(\delta_{\alpha\beta}\delta
(x-y)-\Omega_{\alpha\beta}(x,y))\Psi
   \nonumber\\
& & \
+\frac{1}{2}(u_{\alpha}(x)+\Omega_{\alpha\delta}(x,w)u_{\delta}(w))
(u_{\beta}^{\dagger}(y)-
      u_{\gamma}^{\dagger}(z)\Omega_{\gamma\beta}(z,y))\Psi, \end{eqnarray*}
where a summation (integration) over repeated indices (variables) is
understood.

Using the result (\ref{2.45}) we find eventually for the desired two-point
function the expression
\be \frac{\langle\psi_{\alpha}(x)
\psi_{\beta}^{\dagger}(y)\rangle}{\langle\Psi\vert\Psi\rangle}
=\frac{1}{2}(\delta_{\alpha\beta}\delta(x-y)-
\Omega_{\alpha\beta}(x,y)), \lb{2.46} \ee
or, in operator notation and with respect to a normalized state,
\be \langle\psi(x)\psi^{\dagger}(y)\rangle=\frac{1}{2}(1-\Omega(x,y))=P_+(x,y).
\lb{2.47} \ee
Thus, if one knows the covariance, one can calculate all two-point functions,
and vice versa.
We finally note that excited states can be easily generated by applying the
above creation operator
$a_n^{\dagger}$ on the ground state, leading to a Gaussian times some
polynomial.

\section{Solution of the time-dependent Schr\"odinger equation}

In this section we discuss the solution of the functional Schr\"odinger
equation for fermions in an external electromagnetic field,
\be \left(\int dxdy\psi^{\dagger}(x)h(x,y)\psi(y)\right)\Psi\equiv
H_{\psi}\Psi=i\dot{\Psi}, \lb{2.48} \ee
where, again, $h$ is given explicitly by (\ref{2.5}). Equation (2.48) follows
from a semiclassical expansion of the full functional Schr\"odinger equation
\cite{Ki}. We make again a Gaussian ansatz,
\be \Psi=N(t)\exp\left(u^{\dagger}\Omega(t)u\right), \lb{2.49} \ee
where $\Omega$ and $N$ now depend on time. The state (\ref{2.49}) may be
thought as an evolving vacuum state. Inserting this ansatz into (\ref{2.48}) we
find two equations for $N$ and $\Omega$ which read, in operator notation,
\bea i\frac{d\ln N}{dt} & = & \frac{1}{2}\mbox{Tr}h\Omega \lb{2.50} \\
       i\dot{\Omega} & = & \frac{1}{2}(1-\Omega)h(1+\Omega) \lb{2.51}. \eea
An important special case is given if $\Omega$ can be written in terms of the
projectors (\ref{2.33}, \ref{2.34}). As in the case of the stationary equation
this is equivalent to $\Omega^2=1$.

One physical application we have in mind is to choose the free solution in,
say, the asymptotic past and study its evolution under the influence of an
external electromagnetic field according to (\ref{2.48}). It is important to
note that (\ref{2.51}) preserves the property $\Omega^2=1$. Thus, $\Omega(t)$
can always be written as in (\ref{2.33}) provided $\Omega^2(t_0)=1$  for some
``initial time" $t_0$. This can easily be seen: One first verifies that the
inverse of $\Omega$, $\Omega^{-1}$, obeys the same differential equation as
(\ref{2.51}). From the uniqueness of the solution we thus have
$\Omega(t_0)=\Omega^{-1}(t_0)\Rightarrow\Omega(t)
=\Omega^{-1}(t)\Leftrightarrow\Omega^2(t)=1$.

Eq. (2.51) is solved by
\be \Omega(t) =\left(Q(t)-C\right) \left(Q(t) +C\right)^{-1},
    \lb{2.52} \ee
where $C$ is a time-independent operator, and the operator $Q(t)$ satisfies
\be i\dot{Q}=hQ. \lb{2.53} \ee
One may wish, for example, to choose for $\Omega$ the ``free solution"
(2.32) in the asymptotic past, i.e., one demands that $\Omega$
approaches $\Omega_0=P_- -P_+$ for $t\to-\infty$. This would
correspond to the choice
\bea C &=& P_+ , \lb{2.54} \\
     Q(t)& \stackrel{t\to-\infty}{\longrightarrow}& P_-. \lb{2.55} \eea
The time evolution according to (2.48) will then in general induce
a time dependence of $\Omega$ which may deviate, at late times, from
the asymptotic ``free" solution. This can then be interpreted as particle
creation and will be explicitly discussed below.

The significance of the result (2.52), (2.53) consists in the reduction
of the solution of the full functional equation (2.48) to the solution
of a ``first quantized" problem -- Eq. (2.53) is nothing but the
Dirac equation with an external electromagnetic field.

After the solution for $\Omega$ has been found, the prefactor $N$ can be
immediately determined from (\ref{2.50}) to read
\be N(t)=N_0\exp\left(-\frac{i}{2}\int^t\mbox{Tr}(h\Omega)ds\right). \lb{2.56}
\ee
The time-independent factor $N_0$ can be fixed if $\Psi$ is normalized, i. e.
$\langle\Psi\vert\Psi\rangle=1$, and one finds, using (\ref{2.17}),
\be
N(t)=\mbox{det}^{-1/2}(1+\Omega^{\dagger}\Omega)
\exp\left(-\frac{i}{2}\int^t\mbox{Re\ Tr}(h\Omega)ds
    \right). \lb{2.57} \ee
We now address the question of particle creation. We first note that
 the absolute square of
the matrix element of two Gaussians, $\Psi_1$ and $\Psi_2$, with corresponding
covariances $\Omega_1$ and $\Omega_2$, is given by the expression
\be
\vert\langle\Psi_1\vert\Psi_2\rangle\vert^2=
\mbox{det}\frac{(1+\Omega_1^{\dagger}\Omega_2)
   (1+\Omega_2^{\dagger}\Omega_1)}
     {(1+\Omega_1^{\dagger}\Omega_1)(1+\Omega_2^{\dagger}\Omega_2)}. \lb{2.58}
\ee
In the following we will take for $\Psi_1$ the time-evolved in-vacuum
and for $\Psi_2$ the
vacuum state at late times. The corresponding covariances will be
called $\Omega(t)$ and $\Omega_0$, respectively. As discussed above,
we demand $\Omega(t)$ to approach the ``free covariance" $\Omega_0$
at $t\to-\infty$. Since $\Omega_0=\Omega^{\dagger}$ and
$\Omega_0^2=1$, the desired transition element (2.58) reads
\be
\vert\langle\Psi_1\vert\Psi_2\rangle\vert^2=
\mbox{det}\frac{(1+\Omega_0\Omega(t))
   (1+\Omega^{\dagger}(t)\Omega_0)}
     {2(1+\Omega^{\dagger}(t)\Omega(t))}. \lb{2.59}
\ee
To get the desired expression (2.52) for $\Omega$, which for the
present case reads
\be \Omega(t)= \left(Q(t)-P_+\right)\left(Q(t)+P_+\right)^{-1},
    \lb{2.60} \ee
it is first necessary to solve (2.53) for $Q(t)$. This is most
conveniently done by the ansatz
\be Q(t)= \sum_n \chi_n(t)\chi_n^{\dagger}, \lb{2.61} \ee
where $\chi_n$ (without argument) denotes a negative frequency
eigenfunction of the Dirac Hamiltonian $h$, and $\chi_n(t)$ denotes
the solution of (2.53) which approaches $\chi_n$ in the asymptotic limit
$t\to-\infty$. Therefore,
\[ Q(t)\stackrel{t\to-\infty}{\longrightarrow}
  \sum_n\chi_n\chi_n^{\da}\equiv P_-, \]
as required.

It will prove to be convenient if one expands $\chi_n(t)$ as follows,
\be \chi_n(t)= \alpha_{nm}(t)\chi_m +\beta_{nm}(t)\psi_m, \lb{2.62} \ee
where $\psi_m$ is a positive frequency eigenfunction of $h$,
and $\alpha$, $\beta$ are the time-dependent Bogolubov coefficients
associated with this expansion. Since $h$ is hermitean, the norm
$(\chi_n(t), \chi_m(t))$ is conserved, and we choose it to be
equal to one. The Bogolubov coefficients are then normalized
according to
\be \vert\alpha\vert^2 +\vert\beta\vert^2 =1. \lb{2.63} \ee
Note that this is different from the bosonic case where the analogous
expression contains a minus sign.

The operator $Q(t)+P_+$ in (2.60) is then given by the expression
\be Q(t)+P_+ =\sum_{n,m}\left(\alpha_{nm}\chi_m\chi_n^{\da}
    +\beta_{nm}\psi_m\chi_n^{\da}\right) +\sum_n\psi_n\psi_n^{\da},
    \lb{2.64} \ee
from where its inverse is found to read
\be \left(Q(t)+P_+\right)^{-1} =\sum_n \psi_n\psi_n^{\da}
    -\sum_{n,s,t}\psi_n\alpha_{st}^{-1}\beta_{tn}\chi_n^{\da}
     +\sum_{n,s}\chi_n\alpha_{sn}^{-1}\chi_s^{\da}. \lb{2.65} \ee
One can then write down the desired expression for $\Omega(t)$,
\bea \Omega(t) &=& \sum_n(\chi_n\chi_n^{\da}-\psi_n\psi_n^{\da})
     +2\sum_{n,s,t}\psi_n\alpha_{st}^{-1}\beta_{tn}
     \chi_s^{\da} \nonumber\\
     &=& \Omega_0 +2\sum_{n,s,t}\psi_n\alpha_{st}^{-1}
     \beta_{tn}\chi_s^{\da} \nonumber\\
     &\equiv& \Omega_0+2B, \lb{2.66} \eea
where we have introduced an operator $B$, which in the position
representation is given by
\be B(x,y)=\sum_{n,s,t}\psi_n(x)\alpha_{st}^{-1}\beta_{tn}
    \chi_s^{\da}(y). \lb{2.67} \ee
It maps negative energy eigenfunctions into positive ones, and it
annihilates positive energy eigenfunctions. Conversely, its adjoint
\be B^{\da}(x,y)= \sum_{n,s,t}\chi_s(x)\bar{\alpha}_{st}^{-1}
    \bar{\beta}_{tn}\psi_n^{\da}(y) \lb{2.68} \ee
maps positive energy eigenfunctions into negative ones and
annihilates negative energy eigenfunctions. Note that $B$ and
$B^{\da}$ are nilpotent operators.

One then finds for the various terms in the transition element (2.59)
the expressions
\bea \Omega^{\da}(t)\Omega_0 &=& 1-2B^{\da}, \nonumber\\
     \Omega_0\Omega(t) &=& 1-2B, \nonumber\\
     \Omega^{\da}(t)\Omega(t) &=& 1-2B-2B^{\da}+4B^{\da}B, \lb{2.69} \eea
and one has
\be
\vert\langle\Psi_1\vert\Psi_2\rangle\vert^2=
\mbox{det}\frac{(1-B)
   (1-B^{\da})}
     {(1-B-B^{\da}+2B^{\da}B)}. \lb{2.70}
\ee
Written in the basis $(\psi,\chi)^T$, the various operators in (2.69)
are given by the matrix expressions
\bea B &=& \left( \begin{array}{cc}
      0 & \alpha^{-1}\beta \\
      0 & 0 \end{array} \right), \nonumber\\
     B^{\da} &=& \left( \begin{array}{cc}
      0 & 0 \\
      (\alpha^{-1}\beta)^{\da} &  0 \end{array} \right), \nonumber\\
    B^{\da}B &=& \left( \begin{array}{cc}
      0 & 0 \\
      0 & (\alpha^{-1}\beta)^{\da}
      \alpha^{-1}\beta \end{array} \right). \lb{2.71} \eea
 One immediately verifies that $\mbox{det}(1-B)=
 \mbox{det}(1-B^{\da})=1$. Therefore, using (2.63),
\bea & &
\vert\langle\Psi_1\vert\Psi_2\rangle\vert^2=
\mbox{det}^{-1}(1-B-B^{\da}-2B^{\da}B)
   \nonumber\\
   & & \; = \mbox{det}^{-1}(1+\alpha^{-1}\beta\beta^{\da}
    \alpha^{-1\da}) =\mbox{det}^{-1}(1+\beta^{\da}
    (1-\beta\beta^{\da})^{-1}\beta) \nonumber\\
    & & \ =\mbox{det}^{-1}\beta^{-1}(1-\beta\beta^{\da})^{-1}
    \beta =\mbox{det}(1-\beta\beta^{\da}). \lb{2.72} \eea
The interpretation of this result is obvious. The determinant is less
than one for non-vanishing Bogolubov coefficient $\beta$,
which signals particle creation. Note that the analogous expression
in the bosonic case reads \cite{Ki}
$\mbox{det}^{-1}(1+\beta\beta^{\da})$, which is only equal to
(2.72) for small $\beta$. We will apply the above result to the
calculation of particle creation in an external electric field
in section~5.

  \chapter{Ground state for two-dimensional QED}

\section{The massless case}
\subsection{Calculation of the covariance}

In the following we shall give explicit results for the ground state of
two-dimension\-al QED in arbitrary external electromagnetic fields by applying
the method developed in the last section.
Two-dimensional massless QED is also known as the {\em Schwinger model}
\cite{SM}.
It has been explicitly solved and found to be equivalent to the theory of a
free {\em massive} scalar field (see \cite{SM2} for some literature on the
Schwinger model). In this paper we also address some issues for the Schwinger
model on a finite space \cite{SM3}. The Hamiltonian formalism for the Schwinger
model has been discussed in \cite{Ma} and \cite{Jap}.

It is convenient to discuss the massless and the massive case separately since
it is adequate to use the chiral representation for the Gamma matrices in the
massless case and the Dirac representation in the massive case. For $m=0$ we
thus use
\be \gamma^0= \left( \begin{array}{cc}0 & 1\\ 1 & 0  \end{array} \right), \
      \gamma^1= \left( \begin{array}{cc}0 & 1\\ -1 & 0  \end{array} \right), \
      \gamma^0\gamma^1= \left( \begin{array}{cc}-1 & 0\\ 0 & 1  \end{array}
\right).
      \lb{3.1} \ee
The first-quantized Hamiltonian $h$ (\ref{2.5}) is then given explicitly by
 (with $A^1\equiv A$)
\be h(x,y)=\left(\begin{array}{cc}i\frac{\partial}{\partial x}-eA(x) & 0\\
                                                            0 &
-i\frac{\partial}{\partial x}+eA(x)
                     \end{array}\right) \delta(x-y). \lb{3.2} \ee
To find the ground state of the stationary Schr\"odinger equation we have to
solve the "first-quantized" problem (\ref{2.19}), i. e., to find the spectrum
of (\ref{3.2}),
\be h\psi_n=E_n\psi_n. \lb{3.3} \ee
We quantize the fields in a finite interval, $x\in [0,L]$, and impose the
boundary condition
\be \psi_n(x+L)=e^{2\pi i(\alpha+\beta\gamma_5)}\psi_n(x), \lb{3.4} \ee
where $\alpha$ and $\beta$ are the vectorial and chiral twists, respectively.
Writing
\be \psi_n=\left(\begin{array}{c}\varphi_n\\ \chi_n\end{array}\right), \lb{3.5}
\ee
 the diagonality of $h$ yields two decoupled equations for $\varphi_n$ and
$\chi_n$, corresponding to a decomposition into right- and left handed
fermions.
One finds from (\ref{3.3}) for the right handed part
\bea \varphi_n(x) & = &
\frac{1}{\sqrt{L}}\exp\left[-i\left(E_n^Rx+e\int_0^xA\right)
          \right], \nonumber\\
       E_n^R&=&\frac{2\pi}{L}(n-\alpha-\beta)-\frac{e}{L}\int_0^LA \equiv
\frac{2\pi}{L}(n-\phi), \lb{3.6} \eea
and for the left handed part
\bea \chi_n(x) & = &
\frac{1}{\sqrt{L}}\exp\left[i\left(E_n^Lx-e\int_0^xA\right)
          \right], \nonumber\\
       E_n^L&=&-\frac{2\pi}{L}(n-\alpha+\beta)+\frac{e}{L}\int_0^LA \equiv
-\frac{2\pi}{L}(n-\tilde{\phi}). \lb{3.7} \eea
 Here we have introduced
\bea \phi & = & \alpha+\beta+\frac{e}{2\pi}\int_0^LA \lb{3.8} \\
        \tilde{\phi}& = & \alpha-\beta +\frac{e}{2\pi}\int_0^LA . \lb{3.9} \eea
The covariance (\ref{2.32}) also splits into a right- and left handed part
\be \Omega(x,y)=\left(\begin{array}{cc}\Omega_+(x,y) & 0 \\
                           0 & \Omega_-(x,y)\end{array}\right), \lb{3.10} \ee
where
\bea \Omega_+(x,y) &=& \sum_{E_n^R<0}\varphi_n(x)\varphi^{\dagger}_n(y)
          -   \sum_{E_n^R>0}\varphi_n(x)\varphi^{\dagger}_n(y), \nonumber\\
         \Omega_-(x,y) &=& \sum_{E_n^L<0}\chi_n(x)\chi^{\dagger}_n(y)
          -   \sum_{E_n^L>0}\chi_n(x)\chi^{\dagger}_n(y). \lb{3.11} \eea
{}From (\ref{3.6}) and (\ref{3.7}) one recognizes that $E_n^R>0$ for $n>\phi$
and
$E_n^L>0$ for $n<\tilde{\phi}$. Inserting all this into (\ref{3.11}) one finds
\bea \Omega_+(x,y) &=&
\frac{1}{L}\sum_{E_n^R<0}\exp\left(iE_n^R(y-x)+ie\int_x^yA
       \right) \nonumber\\
       & & \ - \frac{1}{L}\sum_{E_n^R>0}\exp\left(iE_n^R(y-x)+ie\int_x^yA
       \right) \nonumber\\
       &=&
\frac{1}{L}\exp\left(ie\int_x^yA+i\frac{2\pi\phi}{L}(x-y)\right)\times
        \nonumber\\
       & & \ \left(\sum_{n<\phi}\exp\left[-\frac{2\pi in}{L}(x-y)\right]
                   -  \sum_{n>\phi}\exp\left[-\frac{2\pi
in}{L}(x-y)\right]\right)\nonumber\\
     &  = & \frac{i}{L}\exp\left(ie\int_x^yA+\frac{2\pi
i}{L}(\phi-[\phi]-\frac{1}{2})
                   (x-y)\right)\times \nonumber\\  & & \
\frac{1}{\sin\frac{\pi}{L}(x-y)}, \lb{3.12} \eea
where $[\phi]$ denotes the biggest integer smaller or equal than $\phi$.

The left handed part, $\Omega_-(x,y)$, is calculated in the same way, and found
to read
\be \Omega_-(x,y)=-\frac{i}{L}\exp\left(ie\int_x^yA+\frac{2\pi
i}{L}(\tilde{\phi}
-[\tilde{\phi}]-\frac{1}{2})(x-y)\right)\frac{1}{\sin\frac{\pi}{L}(x-y)}.
\lb{3.13} \ee
In the limit $L\to\infty$ the covariance is given by the expression
\be \Omega(x,y)=\frac{i}{\pi}\exp\left(ie\int_x^yA\right){\cal
P}\left(\frac{1}{x-y}\right)
       \left(\begin{array}{cc}1 & 0 \\ 0 & -1 \end{array}\right), \lb{3.14} \ee
where ${\cal P}$ denotes the principal value. This result is in accordance with
\cite{FJ}.
We make a final remark on the existence of {\em large} gauge transformations,
i. e. gauge transformations which cannot be obtained from the identity in a
continuous way. As can be seen from the expressions for the energy, (\ref{3.6})
and (\ref{3.7}), such gauge transformations change the fluxes $\phi$ and
$\tilde\phi$ by an integer. Since the eigenfunctions in (\ref{3.6}) and
(\ref{3.7}) remain unchanged, and the covariance contains only the fractional
part of the flux (see (\ref{3.12}) and (\ref{3.13})),  the wave functional
(\ref{2.14}) remains invariant.

\subsection{Charges and energy}

In this subsection we shall calculate the expectation values of the charge,
chiral charge, and energy with respect to the ground state derived above.

The components of the electric current are given by
\bea j^0 &=&
\psi^{\dagger}\psi=\varphi^{\dagger}\varphi+\chi^{\dagger}\chi\equiv j_++j_-,
\lb{3.15} \\
        j^1 &=&
\psi^{\dagger}\gamma^0\gamma^1\psi=-\varphi^{\dagger}\varphi+
\chi^{\dagger}\chi\equiv
             -j_++j_-. \lb{3.16} \eea
The total charge is thus given by
\be Q=\int dxj_+ + \int dxj_-\equiv Q_+ +Q_-, \lb{3.17} \ee
and the chiral charge by
\be Q_5=Q_+ - Q_-. \lb{3.18} \ee
These expressions contain products of the field operators and thus require a
regularization prescription. The procedure employed here is to first perform a
point splitting and then to subtract the expectation value for vanishing
external field.  After the point splitting is removed, one is left with a
finite result.
The crucial point to note is that the point splitting has to be done in a gauge
invariant way. We thus define the following ``point splitted" quantities
\bea \rho_+(x,y) &=& \varphi^{\dagger}(x)\exp\left(ie\int_x^yA\right)\varphi(y)
\lb{3.19} \\
       \rho_-(x,y)  &=& \chi^{\dagger}(x)exp\left(ie\int_x^yA\right)\chi(y).
\lb{3.20} \eea
They are explicitly gauge invariant.
Applying $\rho_+$ on the vacuum state (\ref{2.14}) we find
\bea \rho_+\Psi &=&
\frac{1}{2}\exp\left(ie\int_x^yA\right)\left(u_1^{\dagger}(x)
+\frac{\delta}{\delta u_1(x)}\right)
\left(u_1(y)+\frac{\delta}{\delta u_1^{\dagger}(y)}\right)\Psi\nonumber\\
 &=&
\frac{1}{2}\exp\left(ie\int_x^yA\right)(\delta(x-y)+
\Omega_+(y,x))\Psi\nonumber\\
  & &
+\frac{1}{2}\exp\left(ie\int_x^yA\right)(u_1^{\dagger}(x)-
\Omega_+(z,x)u_1^{\dagger}(z))\times
    \nonumber\\
   & & \ \ (u_1(y)+\Omega_+(y,z)u(z))\Psi, \lb{3.21} \eea
where, again, an integration over repeated variables is understood. If we set
$x=y$ and integrate over $x$, the last term on the right-hand side of
(\ref{3.21}) vanishes since $(1-\Omega_+)(1+\Omega_+)=0$ according to
(\ref{2.36}). Subtracting the expression for vanishing $A$ field, the first
term after the second equation sign on the right-hand side of (\ref{3.21})
reads
\bea & & \frac{1}{2}\exp\left(ie\int_x^yA\right)\Omega_+(y,x)
-\frac{1}{2}\Omega_{(0)}(y,x) \nonumber\\
& & =\frac{i}{2\pi}  \exp\left(\frac{2\pi
i}{L}(\phi-[\phi]-\frac{1}{2})(y-x)\right)\frac{1}{y-x}
  - \ \phi\leftrightarrow\phi_0 + {\cal O}(x-y), \lb{3.22} \eea
where we have expanded the sine in the expression (\ref{3.12}) for the
covariance and kept only the term proportional to $(x-y)^{-1}$. We have also
introduced
\be \phi_0=\alpha+\beta \lb{3.23} \ee
so that
\be \phi=\phi_0+\frac{e}{2\pi}\int_0^LA\equiv \phi_0+\varphi \lb{3.24} \ee
(compare (\ref{3.8})). Expanding also the exponential in (\ref{3.22}) we note
that the terms which become singular in the limit $x\to y$ drop out. We can
thus remove the point splitting and perform the $x$ integration to find
\be \langle Q_+\rangle=[\phi]-\phi-([\phi_0]-\phi_0). \lb{3.25} \ee
The left handed sector is calculated analogously, with the result
\be \langle
Q_-\rangle=[\tilde{\phi}]-\tilde{\phi}-([\tilde{\phi_0}]-\tilde{\phi_0}),
\lb{3.26} \ee
where
\be \tilde{\phi_0}=\alpha-\beta \lb{3.27} \ee
so that
\be
\tilde{\phi}=\tilde{\phi_0}+\frac{e}{2\pi}\int_0^LA\equiv\tilde{\phi_0}+\varphi
\lb{3.28} \ee
(compare {(3.9)}).

The results for the expectation values of the total charge and chiral charge
are then given by
\bea \langle Q\rangle &=& \langle Q_+\rangle +\langle Q_-\rangle \nonumber\\
 &=&
[\alpha+\beta+\varphi]-[\alpha+\beta]-[\alpha-\beta+\varphi]+[\alpha-\beta]
   \lb{3.29} \eea
and
\bea \langle Q_5\rangle &=& \langle Q_+\rangle -\langle Q_-\rangle \nonumber\\
 &=&
[\alpha+\beta+\varphi]-[\alpha+\beta]+[\alpha-\beta+\varphi]-[\alpha-\beta]
  -2\varphi.  \lb{3.30} \eea
Note that $<Q>=0$ for vanishing chiral twist, $\beta=0$ (see (\ref{3.4})), and
that $<Q_5> =2([\varphi]-\varphi)$ for $\alpha=\beta=0$. The above expectation
values have been calculated, using zeta regularization, by \cite{Jap} for the
special case $\alpha=1/2$ and $\beta=0$. Their result is in agreement with
ours.

We now proceed to calculate the expectation value of the Hamiltonian $H_{\psi}$
(\ref{2.24}).
We first operate with $H_{\psi}$ on the ground state wave functional to find
the expression (\ref{2.27}). We then use the explicit solution (\ref{2.29}) for
the covariance to recognize that only the first term in (\ref{2.27})
contributes to the expectation value $<H_{\psi}>$:
\be \langle H_{\psi}\rangle = \frac{1}{2}\sum_n E_n(1+\Omega_{nn}). \lb{3.31}
\ee
We regularize again by point splitting. We thus introduce a "point splitted"
expectation value which for the contribution from the right handed sector reads
\be \langle\Psi\vert H_{\psi}^{\dagger}(x,y)\vert\Psi\rangle =
\frac{1}{2}\exp\left(-ie\int_x^yA\right)h_x
\sum_n(1+\Omega_{nn})\varphi_n(x)\varphi_n^{\dagger}(y).
 \lb{3.32} \ee
Note that this expression is explicitly gauge-invariant and reduces to
(\ref{3.31}) after setting $x=y$ and integrating over $x$ (the action of the
first-quantized Hamiltonian $h_x\equiv i\partial/\partial x - eA(x)$ just
produces the energy $E_n$ when acting on the $\psi_n$). The completeness of the
$\varphi_n$, as well as (\ref{2.26}), enables one to write (\ref{3.32}) as
\be \langle\Psi\vert H_{\psi}^{\dagger}(x,y)\vert\Psi\rangle =
\frac{1}{2}\exp\left(-ie\int_x^yA\right)h_x
     (\delta(x-y)+\Omega_+(x,y)). \lb{3.33} \ee
Using the explicit expression (\ref{3.12}) for $\Omega_+(x,y)$ one finds, up to
order $x-y$,
\begin{eqnarray*} & & \exp\left(-ie\int_x^yA\right)h_x\Omega_+(x,y)=\\
   & & \
\left(-\frac{2i}{L(x-y)}(\phi-[\phi]-\frac{1}{2})
+\frac{1}{\pi(x-y)^2}-\frac{\pi}{6L^2}\right)
\times\\ & & \ \
 \exp\left(\frac{2\pi i}{L}(\phi-[\phi]-\frac{1}{2})(x-y)\right) +{\cal
O}(x-y).
\end{eqnarray*}
Expanding also the exponential, this reads
\[\exp\left(-ie\int_x^yA\right)h_x\Omega_+(x,y) =
\frac{1}{\pi(x-y)^2}-\frac{\pi}{6L^2}
   +\frac{2\pi}{L^2}(\phi-[\phi]-\frac{1}{2})^2+{\cal O}(x-y), \]
so that we find
\begin{eqnarray*} \langle\Psi\vert H_{\psi}^{\dagger}(x,y)\vert\Psi\rangle &=&
\frac{1}{2}\exp\left(-ie\int_x^yA\right)
      (i\frac{\partial}{\partial x}-eA)\delta(x-y)\\ & & \
+\frac{1}{2\pi(x-y)^2}-\frac{\pi}{12L^2}
   +\frac{\pi}{L^2}(\phi-[\phi]-\frac{1}{2})^2+{\cal O}(x-y). \end{eqnarray*}
Since
\[ \exp\left(-ie\int_x^yA\right)
      i\frac{\partial}{\partial x}\delta(x-y)=i\delta'(x-y)+eA(x)\delta(x-y),
\]
we have
\be \langle\Psi\vert H_{\psi}^{\dagger}(x,y)\vert\Psi\rangle =
\frac{i}{2}\delta'(x-y)
   +\frac{1}{2\pi(x-y)^2}-\frac{\pi}{12L^2}
   +\frac{\pi}{L^2}(\phi-[\phi]-\frac{1}{2})^2+{\cal O}(x-y). \lb{3.34} \ee
{}From this expression one has to subtract the expectation value for vanishing
external field. To retain finite-size effects we subtract the "free" value for
$L\to\infty$. This removes the divergent terms in (\ref{3.34}). Setting $x=y$
and integrating over $x$, one finds the result
\be \langle H_{\psi}^{\dagger}\rangle =
\frac{\pi}{L}\left(\phi-[\phi]-\frac{1}{2}\right)^2
       -\frac{\pi}{12L}. \lb{3.35} \ee
This vanishes in the limit $L\to\infty$. The expression for finite $L$ is
nothing but the {\em Casimir energy} which is also present for vanishing
external field:
\[ \langle H_{\psi}^{\dagger}\rangle =
\frac{\pi}{L}\left(\phi_0-[\phi_0]-\frac{1}{2}\right)^2
       -\frac{\pi}{12L}.  \]
Note that the resulting force between the boundaries at $x=0$ and $x=L$ can be
attractive or
repulsive, depending on the chosen boundary conditions. For the conditions
chosen in \cite{Jap} the expectation value is given by $-\pi/12L$ and thus
leads to an attractive force.

The expectation value of the Hamiltonian in the left handed sector is
calculated in the same way by making use of (\ref{3.13}) and using
$-h_x=-i\partial/\partial x +eA(x)$. Instead of (\ref{3.35}) one finds
\be \langle H_{\psi}^-\rangle =
\frac{\pi}{L}\left(\tilde{\phi}-[\tilde{\phi}]-\frac{1}{2}\right)^2
       -\frac{\pi}{12L}. \lb{3.36} \ee
The total Casimir energy is the sum of the expressions (\ref{3.35}) and
(\ref{3.36}).

\section{The massive case}
\subsection{Calculation of the covariance}

In the massive case we use the Dirac representation for the Gamma matrices, i.
e.,
\be \gamma^0= \left( \begin{array}{cc}1 & 0\\ 0 & -1  \end{array} \right), \
      \gamma^1= \left( \begin{array}{cc}0 & -1\\ 1 & 0  \end{array} \right), \
      \gamma^0\gamma^1= \left( \begin{array}{cc}0 & -1\\ -1 & 0  \end{array}
\right).
      \lb{3.37} \ee
The first-quantized Hamiltonian is then given by the expression
\be h(x,y)=\left(\begin{array}{cc} m & i\frac{\partial}{\partial x}-eA(x) \\
i\frac{\partial}{\partial x}-eA(x)& -m
                     \end{array}\right) \delta(x-y). \lb{3.38} \ee
We are again looking for the eigenfunctions of $h$,
\be h\psi_n=E_n\psi_n. \lb{3.39} \ee
If we make the ansatz
\be \psi_n=\frac{1}{\sqrt{L}}\exp\left(-ie\int_0^xA-i\lambda_nx\right)c_n,
\lb{3.40} \ee
Eq. (\ref{3.39}) yields an algebraic equation for $c_n$,
\be \left(\begin{array}{cc}m-E_n & \lambda_n\\ \lambda_n & -m-E_n
\end{array}\right)
      \left(\begin{array}{c}c_{n,1}\\c_{n,2}\end{array}\right)=0. \lb{3.41} \ee
The boundary condition
\be \psi_n(x+L)=e^{2\pi i\alpha}\psi_n(x) \lb{3.42} \ee
yields a quantization condition for the $\lambda_n$,
\be
\lambda_n=\frac{2\pi}{L}\left(n-\alpha-\frac{e}{2\pi}
\int_0^LA\right)\equiv\frac{2\pi}{L}
      (n-\phi), \lb{3.43} \ee
where $n\in Z$.
{}From (\ref{3.41}) one then finds the values for the energy,
\be
E_n=\pm\sqrt{m^2+\lambda_n^2}=\pm\sqrt{m^2+
\frac{4\pi^2}{L^2}(n-\phi)^2}\equiv
      \pm\omega_n. \lb{3.44} \ee
We already note at this point that the massless limit of (\ref{3.44}) yields
$E_n=\pm\frac{2\pi}{L}
 \vert n-\phi\vert$ instead of $E_n=\pm\frac{2\pi}{L}
 ( n-\phi)$ which was found by starting from $m=0$ ab initio. This will be
relevant for the discussion of anomalies in chapter~5.

The normalized eigenfunctions $\psi_n$ read
\be \psi_{n,+}=\frac{1}{\sqrt{2\omega_n(\omega_n+m)L}}\left(\begin{array}{c}
       \omega_n+m\\ \lambda_n
\end{array}\right)\exp\left(-i\lambda_nx-ie\int_0^xA\right)
       \lb{3.45} \ee
for $E_n=\omega_n$, and
 \be
\psi_{n,-}=\frac{1}{\sqrt{2\omega_n(\omega_n+m)L}}
\left(\begin{array}{c}-\lambda_n\\
       \omega_n+m\end{array}\right)\exp\left(-i\lambda_nx-ie\int_0^xA\right)
       \lb{3.46} \ee
for $E_n=-\omega_n$.

We now use again (\ref{2.32}) and the filling prescription (\ref{2.31}) to
calculate the covariance,
\bea \Omega(x,y) &=&
\sum_n\psi_{n,-}(x)\psi^{\dagger}_{n,-}(y)
-\sum_n\psi_{n,+}(x)\psi^{\dagger}_{n,+}(y)
     \nonumber\\
     &\equiv& P_--P_+. \lb{3.47} \eea
Noting that $\lambda_n=(\omega_n+m)(\omega_n-m)$, we find
\be
P_+(x,y)=\frac{1}{2L}\exp\left(-ie\int_y^xA\right)
\sum_n\frac{e^{-i\lambda_n(x-y)}}{\omega_n}
 \left(\begin{array}{cc}\omega_n+m & \lambda_n\\ \lambda_n & \omega_n-m
 \end{array}\right) \lb{3.48} \ee
and
\be
P_-(x,y)=\frac{1}{2L}\exp\left(-ie\int_y^xA\right)
\sum_n\frac{e^{-i\lambda_n(x-y)}}{\omega_n}
 \left(\begin{array}{cc}\omega_n-m & -\lambda_n\\ -\lambda_n & \omega_n+m
 \end{array}\right) . \lb{3.49} \ee
To evaluate the various sums in these expressions we make use of Poisson's
summation formula:
\be 2\pi\sum_{n=-\infty}^{\infty}f(2\pi n)=\sum_{n=-\infty}^{\infty}F(n),
\lb{3.50} \ee
where
\be F(u)=\int_{-\infty}^{\infty}dzf(z)e^{izu}. \lb{3.51} \ee
We then have
\be \sum_n\frac{\lambda_n}{\omega_n}e^{-2\pi in(x-y)/L}\equiv
      \sum_n f(2\pi n) \lb{3.52} \ee
and
\be \sum_n\frac{1}{\omega_n}e^{-2\pi in(x-y)/L}\equiv
      \sum_n \tilde{f}(2\pi n), \lb{3.53} \ee
where
\be f(z)=\frac{z-2\pi\phi}{\sqrt{(z-2\pi\phi)^2+m^2L^2}}e^{-i(x-y)z/L}
\lb{3.54} \ee
and
\be \tilde{f}(z)=\frac{L}{\sqrt{(z-2\pi\phi)^2+m^2L^2}}e^{-i(x-y)z/L}.
\lb{3.55} \ee
{}From (\ref{3.51}) we then find
\bea F(u) & = & e^{2\pi
i\phi(u-(x-y)/L)}\int_{-\infty}^{\infty}dp\frac{pe^{ip(u-(x-y)/L)}}
       {\sqrt{p^2+m^2L^2}} \nonumber\\
     & = & 2ie^{2\pi i\phi(u-(x-y)/L}mLK_1(mLu-m(x-y)) \lb{3.56} \eea
and
\bea \tilde{F}(u) & = &L e^{2\pi
i\phi(u-(x-y)/L)}\int_{-\infty}^{\infty}dp\frac{e^{ip(u-(x-y)/L)}}
       {\sqrt{p^2+m^2L^2}} \nonumber\\
     & = & 2Le^{2\pi i\phi(u-(x-y)/L}K_0(mLu-m(x-y)). \lb{3.57} \eea
Here $K_0$ and $K_1$ denote Bessel functions and use has been made of \cite{GR}
to evaluate the integrals. From (\ref{3.50}) we then find for the sums
\be \sum_n\frac{\lambda_n}{\omega_n}e^{-2\pi
in(x-y)/L}=-\frac{imL}{\pi}e^{-2\pi i\phi(x-y)/L}\sum_n e^{-2\pi
in\phi}K_1(mLn+m(x-y)) \lb{3.58} \ee
and
\be \sum_n\frac{1}{\omega_n}e^{-2\pi in(x-y)/L}=\frac{L}{\pi}e^{-2\pi
i\phi(x-y)/L}\sum_n e^{-2\pi in\phi}K_0(mLn+m(x-y)). \lb{3.59} \ee
In the expressions below we will for simplicity not explicitly write out the
argument
$mLn+m(x-y)$ of the Bessel functions $K_0$ and $K_1$.
For the remaining sum we have
 \be \sum_ne^{-\frac{2\pi in}{L}(x-y)}=L\sum_n\delta(x-y-nL)=L\delta(x-y)
\lb{3.60} \ee
since $\vert x-y\vert<L$.
Inserting all these results into the expressions (\ref{3.48}) and (\ref{3.49})
we find
\bea & &  P_+ = \frac{1}{2}\delta(x-y){\bf I}  +
\frac{m}{2\pi}\exp\left(-ie\int_y^xA\right)\times\nonumber\\ & & \
\sum_n\left(\begin{array}{cc}
      K_0 & -iK_1\\-iK_1 & -K_0
     \end{array} \right)e^{-2\pi in\phi}   \lb{3.61} \eea
and
\bea & & P_- =  \frac{1}{2}\delta(x-y){\bf I}
-\frac{m}{2\pi}\exp\left(-ie\int_y^xA\right)\times\nonumber\\ & &
\sum_n\left(\begin{array}{cc}
      K_0 & -iK_1\\-iK_1 & -K_0
     \end{array} \right)e^{-2\pi in\phi}  . \lb{3.62} \eea
We verify that $P_+ +P_-={\bf I}$. Our final result for the covariance
(\ref{2.33}) is then given by the expression
\bea & & \Omega(x,y)=P_--P_+=\frac{m}{\pi}\exp\left(-ie\int_y^xA\right)\times
\nonumber\\& & \
\sum_n\left(\begin{array}{cc}
      -K_0 & iK_1\\iK_1 & K_0
     \end{array} \right)e^{-2\pi in\phi}. \lb{3.63} \eea
In the limit $L\to\infty$ we find
\be
\Omega(x,y)=\frac{m}{\pi}\exp\left(-ie\int_y^xA\right)\left(\begin{array}{cc}
      -K_0(m(x-y)) & iK_1(m(x-y))\\iK_1(m(x-y)) & K_0(m(x-y))
     \end{array} \right) . \lb{3.64} \ee
Using the asymptotic expressions for the Bessel functions one verifies that
$\Omega$ approaches the result (\ref{3.14}) in the limit of vanishing mass.
Furthermore, in the opposite limit of large mass (or large $\vert x-y\vert$)
the covariance reads
\be \Omega(x,y)=\sqrt{\frac{m}{2\pi\vert
x-y\vert}}\exp\left(-ie\int_y^xA\right)
    \exp\left(-m\vert x-y\vert\right)\left(\begin{array}{cc}-1 & i\\ i & 1
\end{array}\right).
     \lb{3.65} \ee
In summary, we have found in this section the exact ground state for arbitrary
external fields in the massive Schwinger model. The excited states,
$\Psi_n$, can then be constructed in the usual way through the
application of the creation operator. A general state, $\Psi$,
of the fully quantized theory can then be expanded into these
energy eigenstates according to
\[ \Psi[A,u,u^{\da}] =\sum_n \varphi_n[A]\Psi_n[A,u,u^{\da}], \]
where the functionals $\varphi_n[A]$ can be determined from the
full functional Schr\"odinger equation which contains the kinetic term
$-\delta^2/2\delta A^2$ in the Hamiltonian.

\subsection{Charges and energy}

We define again a "point splitted" charge operator
\be \rho(x,y)=\psi^{\dagger}(x)\exp\left(ie\int_x^yA\right)\psi(y) \lb{3.66}
\ee
and find for its action on the vacuum state an expression analogous to
(\ref{3.21})
(there is now no distinction between a left and a right handed sector):
\bea \rho(x,y)\Psi  &=&
\frac{1}{2}\exp\left(ie\int_x^yA\right)(2\delta(x-y)
+\sum_{\alpha=1}^2\Omega_{\alpha\alpha}
(y,x))\Psi
\nonumber\\
  & & +
\frac{1}{2}\exp\left(ie\int_x^yA\right)
(u_{\alpha}^{\dagger}(x)-\Omega_{\beta\alpha}(z,x)u_{\beta}
^{\dagger}(z))\times
    \nonumber\\
   & & \  (u_{\alpha}(y)+\Omega_{\alpha\gamma}(y,z)u_{\gamma}(z))\Psi,
\lb{3.67} \eea
where a summation (integration) over repeated indices (variables) is
understood.
Like in the massless case, the second term on the right-hand side of
(\ref{3.67}) vanishes after setting $x=y$ and integrating over $x$. The first
term is again regularized by subtracting its value for vanishing external
field. This yields for the vacuum expectation value of the total charge
\be \langle Q\rangle=\frac{1}{2}\sum_{\alpha=1}^2\int_0^L dx\lim_{x\to y}\left(
     \Omega_{\alpha\alpha}(y,x)-\Omega_{\alpha\alpha}^{(0)}(y,x)\right)=0,
\lb{3.68} \ee
since the covariance (\ref{3.63}) is traceless with respect to the spinor
indices. The result (\ref{3.68}) has of course been expected since the total
charge should annihilate the vacuum state (see also the discussion in
section~6). This is true in any number of dimensions.

For the chiral charge we give first a general expression which is valid in any
even dimension. We define the ``point splitted" chiral charge
\bea \rho_5(x,y)&=&
\bar{\psi}(x)\gamma^5\gamma^0\exp\left(ie\int_x^yA\right)\psi(y)
     \nonumber\\
  &=& - \psi^{\dagger}(x)\gamma^5\exp\left(ie\int_x^yA\right)\psi(y). \lb{3.69}
\eea
Operating with this on the vacuum state yields (compare (\ref{3.21}))
\bea \rho_5(x,y)\Psi  &=& -\frac{1}{2}\exp\left(ie\int_x^yA\right)\mbox
{Tr}\gamma^5
(\delta(x-y)+\Omega
(y,x))\Psi
\nonumber\\
  & & - \frac{1}{2}\exp\left(ie\int_x^yA\right)\int dvdw
u^{\dagger}(v)\left(\delta(v-x)
-\Omega(v,x)\right)\gamma^5
 \times    \nonumber\\
   & & \  \left(\delta(y-w)+\Omega(y,w)\right)u(z)\Psi. \lb{3.70} \eea
The second term can be written, after setting $x=y$, integrating over $x$, and
performing the expectation value, as
\[ -2\mbox {Tr}P_+\gamma^5P_-=0 \]
since $P_+P_-=0$ (compare (\ref{3.35})), and use has been made of (\ref{2.45}).
We are thus left with
\be \langle\Psi\vert Q_5\vert\Psi\rangle = -\frac{1}{2}\mbox{Tr}\int_0^L
dx\lim_{x\to y}
       \gamma^5\Omega(y,x)\exp\left(ie\int_x^yA\right), \lb{3.71} \ee
from where the result for $A=0$ has to be subtracted. Using the explicit
results in two dimensions we find
\begin{eqnarray*} & &  -\frac{1}{2}\mbox{ Tr }
       \gamma^5\Omega(y,x)\exp\left(ie\int_x^yA\right) \\
      & & \ =\frac{im}{\pi}\sum_nK_1(mLn+m(y-x))e^{-2\pi in\phi}.
\end{eqnarray*}
Subtracting from this the expression with $A=0$ we get
\[ \frac{im}{\pi}\sum_n\left( K_1(m(y-x)+nmL)e^{-2\pi
in\phi}-K_1(m(y-x)+nmL)e^{-2\pi in\alpha}\right) \]
so that we have
\bea & & \langle Q_5\rangle =\frac{im}{\pi}\int_0^L dx\sum_{n\neq 0}K_1(nmL)
\left( e^{-2\pi in\phi}
       -e^{-2\pi in\alpha}\right)\nonumber\\
 & & \ =\frac{2mL}{\pi}\sum_{n>0}K_1(nmL)(\sin(2\pi n\phi)-\sin(2\pi n\alpha)).
\lb{3.72} \eea
In the limit $m\to 0$ we obtain
\bea \lim_{m\to 0}\langle Q_5\rangle &=& \frac{2}{\pi}\sum_{n>0}\frac{1}{n}
    (\sin(2\pi n\phi)-\sin(2\pi n\alpha))\nonumber\\
     &=& 2([\phi]-\phi+\frac{1}{2})-2([\alpha]-\alpha+\frac{1}{2})\nonumber\\
    &=& 2([\alpha+\varphi]+[\alpha]-\varphi). \lb{3.73} \eea
This is equal to our earlier result (\ref{3.30}) when evaluated for $\beta=0$
($\varphi$ was defined in (\ref{3.28})).
Recalling the asymptotic formula for $K_1$ in the limit of large arguments one
finds that
\be \langle Q_5\rangle \stackrel{L\to\infty}{\sim}\sqrt{\frac{2mL}{\pi}}
      (\sin(2\pi n\phi)-\sin(2\pi n\alpha))e^{-mL}\stackrel{L\to\infty}{\to}0.
\lb{3.74} \ee
We finally calculate the vacuum expectation value of the Hamiltonian $H_{\psi}$
 (\ref{2.24}) in the massive case. We start from the expectation value
(\ref{3.33}) for the point splitted Hamiltonian but insert in that expression
\be h_x=-i\gamma^0\gamma^1(\partial_x-iA)+m\gamma^0 \lb{3.75} \ee
 as well as the full covariance $\Omega$ instead of $\Omega_+$. With our result
(\ref{3.63}) for the covariance we then find
\begin{eqnarray*}& & \exp\left(ie\int_y^xA\right)
 h_x\Omega(x,y) = -\frac{im^2}{\pi}
\gamma^0\gamma^1
\sum_n\left(\begin{array}{cc}
      -K_0' & iK_1'\\iK_1' & K_0'
     \end{array} \right)e^{-2\pi in\phi}\\
& & \ \ +\frac{m^2}{\pi}\gamma^0\exp\left(-ie\int_y^xA\right)
     \sum_n\left(\begin{array}{cc}
      -K_0 & iK_1\\iK_1 & K_0
     \end{array} \right)e^{-2\pi in\phi}. \end{eqnarray*}
Thus,
\bea & & \langle\Psi\vert H_{\psi}(x,y)\vert\Psi\rangle =
\frac{1}{2}\mbox{Tr}\exp\left(-ie\int_y^xA\right)
(h_x\Omega(x,y)+h_x\delta(x-y))
      \nonumber\\
    & & \ = \mbox{Tr}\left\{\frac{m^2}{2\pi}
       \sum_n\left(\begin{array}{cc}
      -K_0-K_1' & iK_1+iK_0'\\-iK_1 -iK_0'& -K_1'-K_0
     \end{array} \right)e^{-2\pi in\phi}\right.\nonumber\\
  & & \  \left. +\frac{1}{2}\left(\begin{array}{cc}0&1\\1&0\end{array}\right)
i\delta'(x-y)+\frac{m}{2}\left(\begin{array}{cc}1&0\\0&-1\end{array}\right)
     \delta(x-y)\right\}. \lb{3.76} \eea
{}From this we subtract the expectation value for $L\to\infty$ and vanishing
$A$.
Using the relations
\[ K_1'(\xi)+K_0(\xi)=-\frac{K_1(\xi)}{\xi}, \ \  K_1=-K_0', \]
 this yields
\bea \langle H_{\psi}\rangle &=& \int_0^L dx\lim_{x\to y}\left(\langle\Psi\vert
 H_{\psi}(x,y)\vert\Psi\rangle -\langle\Psi_0\vert
H_{\psi}(x,y)\vert\Psi_0\rangle \right)
  \nonumber\\
  &=& \frac{m^2}{\pi} \int_0^L dx\lim_{x\to y}
\left(\sum_n\frac{e^{-2\pi in\phi}}
    {m(x-y)}K_1(m(x-y))\right.\nonumber\\
 & & \  \left.-\frac{1}
    {m(x-y)+nmL}K_1(m(x-y)+nmL)\right) \nonumber\\
&=& \frac{m}{\pi}\sum_{n\neq 0}\frac{1}{n}K_1(nmL)e^{-2\pi in\phi}\nonumber\\
& = & \frac{2m}{\pi}\sum_{n>0}\frac{1}{n}K_1(nmL)\cos(2\pi n\phi). \lb{3.77}
\eea
This vanishes in the limit $L\to\infty$ but remains finite for finite $L$ even
for vanishing electromagnetic field where we have
 \be \langle H_{\psi}\rangle_{A=0}=\frac{2m}{\pi}\sum_{n>0}\frac{1}{n}K_1(nmL)
\cos(2\pi n \alpha). \lb{3.78} \ee
In the limit of vanishing mass we obtain from (\ref{3.77}) the result of
section~3.1.2.
The expectation value of the Hamiltonian vanishes for $L\to\infty$ as can be
easily seen from (\ref{3.77}).

\chapter{Non-abelian gauge fields}

\section{Calculation of the covariance}

We consider the Lagarangian
\be {\cal L}=-\frac{1}{4}F_{\mu\nu}F^{\mu\nu}
+\bar{\psi}(iD_{\mu}\gamma^{\mu}-m)\psi, \lb{4.1} \ee
where
\bea F_{\mu\nu}&=& \partial_{\mu}A_{\nu}-\partial_{\nu}A_{\mu} +i[A_{\mu},
A_{\nu}] \lb{4.2} \\
  D_{\mu}&=&\partial_{\mu}+iA_{\mu}. \lb{4.3} \eea
We have introduced here the matrix-valued vector field, $A_{\mu}(x)$, which is
defined by
\be A_{\mu}=A_{\mu}^iT_i, \lb{4.4} \ee
where $T_i$ are the (hermitean) generators of a Lie group, which are normalized
according to
\be (T_i,T_j)=\delta_{ij}.  \lb{4.5} \ee
The gauge coupling constant has been set equal to one. In two dimensions the
discussion is greatly simplified since the gauge $A^0=0$ removes the commutator
in (\ref{4.2}). This enables us to proceed analogously to the abelian case.
Denoting $A_1\equiv A$, the total Hamiltonian density reads explicitly
\bea {\cal H} &=&
\frac{1}{2}\pi_A^2-i\psi^{\dagger}\gamma^0\gamma^1(\partial_x+iA)\psi
+m\psi^{\dagger}\gamma^0\psi \nonumber\\
  &\equiv&  \frac{1}{2}\pi_A^2+\psi^{\dagger}h\psi. \lb{4.6} \eea
The first-quantized Hamiltonian reads
\be h=gh_{(0)}g^{-1}, \lb{4.7} \ee
where
\be h_{(0)}=-i\gamma^0\gamma^1\partial_x +m\gamma^0 \lb{4.8} \ee
and
\be g(x)={\cal P}\exp\left(-i\int_0^xA\right), \lb{4.9} \ee
where ${\cal P}$ denotes path-ordering (we will suppress this letter in the
following).
It follows immediately that if $\psi^{(0)}$ is an eigenfunction of $h_{(0)}$
with eigenvalue $E$ then $\psi=g\psi^{(0)}$ is an eigenfunction of $h$ with the
same eigenvalue $E$. From (\ref{3.45}) and (\ref{3.46}) we see that the free
eigenfunctions are given by
\be \psi^{(0)}_{n,+}=cf_n, \ \  \psi^{(0)}_{n,-}=cg_n,  \lb{4.10} \ee
where $c$ is a constant vector in the representation space of the above
generators, and
\be f_{n}(x)=\frac{1}{\sqrt{2\omega_n(\omega_n+m)L}}\left(\begin{array}{c}
       \omega_n+m\\ \lambda_n \end{array}\right)\exp\left(-i\lambda_nx\right)
       \lb{4.11} \ee
 and
 \be g_{n}(x)=\frac{1}{\sqrt{2\omega_n(\omega_n+m)L}}\left(\begin{array}{c}
-\lambda_n\\
       \omega_n+m\end{array}\right)\exp\left(-i\lambda_nx\right) . \lb{4.12}
\ee
We also have to implement the boundary conditions
\be \psi(L)=e^{2\pi i\alpha}\psi(0)=g(L)\psi^{(0)}(L)=g(0)e^{2\pi i\alpha}
\psi^{(0)}(0). \lb{4.13} \ee
We note that
\be g^{-1}(L)g(0)=\exp\left(i\int_0^LA\right)\equiv\exp\left(iB\right).
\lb{4.14} \ee
Since $B$ is a hermitean matrix it can be diagonalized:
\bea Be_a &=& \mu_ae_a, \lb{4.15} \\
        (e_a,e_b) &=& \delta_{ab}, \lb{4.16} \eea
where $a$ and $b$ run form $1$ to the dimension of the representation.
We thus have
\be  g^{-1}(L)g(0)e_a=\exp\left(i\mu_a\right)e_a. \lb{4.17} \ee
Choosing $c=e_a$ we find from (\ref{4.13}) the quantization condition
\be \lambda_{n,a}=\frac{2\pi}{L}(n-\alpha)-\frac{\mu_a}{L}, \lb{4.18} \ee
and the energies are given by
\be E_{n,a}=\pm\sqrt{m^2+\lambda^2_{n,a}}\equiv\pm\omega_{n,a} \lb{4.19} \ee
in analogy to the abelian result (\ref{3.44}).
{}From (\ref{4.10}) and $\psi=g\psi^{(0)}$ the positive energy solutions are
given by
\be \psi^a_{n,+}=g(x)e_a\otimes f_n^a, \lb{4.20} \ee
and the negative energy solutions by
\be \psi^a_{n,-}=g(x)e_a\otimes g_n^a, \lb{4.21} \ee
(no summation over $a$).
These solutions are orthonormal since
\be (\psi^a_{n,+},\psi^b_{m,+})=(g(x)e_a\otimes f_n^a,g(x)e_b\otimes f_m^b)
     =\delta_{ab}\delta_{nm}, \ \ \mbox{etc.}  \lb{4.22} \ee
Under a gauge transformation mediated by $U(x)$ the following transformation
laws hold:
\bea \psi &\to & \tilde{\psi}=U(x)\psi, \lb{4.23} \\
        A &\to & \tilde{A}=UAU^{-1}+i(\partial_xU)U^{-1}, \lb{4.24} \\
      g &\to & \tilde{g}=U(x)g(x)U^{-1}(0), \lb{4.25} \\
   \psi^{(0)} &\to & \tilde{\psi}^{(0)}=U(0)\psi^{(0)}. \lb{4.26} \eea
Since gauge transformations should respect the boundary conditions, we must
have $U(0)=U(L)$. Since the "boundary operator" $g^{-1}(L)g(0)$ transforms as
\[ g^{-1}(L)g(0) \to U(0)g^{-1}(L)g(0)U^{-1}(0) \]
the quantities $\mu_a$ appearing in (\ref{4.17}) are gauge invariant.

We now proceed to calculate the covariance of the ground state.
For the projector on positive energies one finds, making use of the result
(\ref{3.48}) for the abelian case,
\bea P_+(x,y) &=& \sum_{a,n}\psi^a_{n,+}(x)\psi^{a+}_{n,+}(y) \nonumber\\
 &=& \frac{1}{2L}g(x)\left[\sum_a e_ae_a^{\dagger}\exp\left(\frac{i}{L}
  (2\pi\alpha+\mu_a)(x-y)\right)\times \right.\nonumber\\
   & & \  \left. \sum_n\frac{e^{-2\pi in(x-y)/L}}{\omega_n}
    \left(\begin{array}{cc}\omega_n+m &\lambda_n \\ \lambda_n & \omega_n-m
      \end{array}\right)\right]g^{\dagger}(y). \lb{4.27} \eea
Applying Poisson's summation formula (\ref{3.50}) one finds, in analogy to
(\ref{3.61}),
\bea   P_+(x,y) &=& \frac{1}{2}\delta(x-y){\bf I}  -
\frac{m}{2\pi}g(x)\left[\sum_{a,n}e_a
\exp\left(-in\int_0^L\mu_a\right)e_a^{\dagger}\times
\right.\nonumber\\ & & \ \left.
\left(\begin{array}{cc}
     - K_0 & iK_1\\iK_1 &  K_0
     \end{array} \right)e^{-2\pi in\alpha}\right] g^{\dagger}(y).  \lb{4.28}
\eea
and
\be  P_-(x,y) = \delta(x-y){\bf I}-P_+(x,y).  \lb{4.29} \ee
It is convenient to define the "diagonal matrix"
\be D=\sum_a\exp\left(i\mu_a\right)e_ae_a^{\dagger} \lb{4.30} \ee
which obeys
\be D^n=\sum_a\exp\left(in\mu_a\right)e_ae_a^{\dagger}. \lb{4.31} \ee
The covariance can thus be written as
\be  \Omega(x,y)=P_--P_+=\frac{m}{\pi}g(x)\left[
\sum_n D^{-n}e^{-2\pi in\alpha}
\left(\begin{array}{cc}
      -K_0 & iK_1\\iK_1 & K_0
     \end{array} \right) \right] g^{\dagger}(y). \lb{4.32} \ee
In the abelian case we have
\be \mu=\int_0^LA \lb{4.33} \ee
so that the result (\ref{4.32}) equals our earlier result (\ref{3.63}).

\section{Charges and energy}

The point splitted version of the non-abelian current operator reads, in any
number of dimensions,
\be
j_i^{\mu}(x,y)=\psi^{\dagger}(x)\exp\left(i\int_x^yA\right)
T_i\gamma^0\gamma^{\mu}
\psi(y). \lb{4.34} \ee
Its action on the vacuum state $\Psi$ can be found in the same way as for the
abelian case (\ref{3.67}). The result is
 \bea j^{\mu}_i(x,y)\Psi &=& \frac{1}{2}\mbox{Tr}\exp\left(i\int_x^yA\right)
  T_i\gamma^0\gamma^{\mu}(\delta(x-y)+\Omega(y,x))\Psi\nonumber\\
& & +\frac{1}{2}\int dvdw
u^{\dagger}(v)(\delta(v-x)-\Omega(v,x))
\exp\left(i\int_x^yA\right)\times\nonumber\\
 & & \  \ T_i\gamma^0\gamma^{\mu}(\delta(y-w)+\Omega(y,w))u(w)\Psi.
  \lb{4.35} \eea
Taking the expectation value of the second term in (\ref{4.35}) with respect to
$\Psi$, one gets, making use of (\ref{2.45}) and (\ref{2.35})
\bea & &
\frac{1}{4}\mbox{Tr}\int dvdw\
(\delta(v-x)-\Omega(v,x))\exp\left(i\int_x^yA\right)\times\nonumber\\
 & & \ \
T_i\gamma^0\gamma^{\mu}(\delta(y-w)+\Omega(y,w))\Omega(w,v)\nonumber\\
&=& \frac{1}{4}\mbox{Tr}\int
dv(\delta(y-v)+\Omega(y,v))(\delta(v-x)-\Omega(v,x))\times\nonumber\\
& &  \ \exp\left(i\int_x^yA\right)T_i\gamma^0\gamma^{\mu}=0. \lb{4.36} \eea
The expectation value of the point splitted current with respect to $\Psi$ is
thus given by
\be \langle\Psi\vert j_i^{\mu}(x,y)\vert\Psi\rangle=\frac{1}{2}\mbox{Tr}
 \exp\left(i\int_x^yA\right)T_i\gamma^0\gamma^{\mu}(\delta(x-y)+\Omega(y,x)).
\lb{4.37} \ee
For the axial current
\be j_{5i}^{\mu}(x,y)=\psi^{\dagger}(x)\exp\left(i\int_x^yA\right)T_i\gamma^0
\gamma^5\gamma^{\mu}
\psi(y) \lb{4.38} \ee
the analogous result is (compare also (\ref{3.70}))
\be \langle\Psi\vert j_{5i}^{\mu}(x,y)\vert\Psi\rangle=-\frac{1}{2}\mbox{Tr}
 \exp\left(i\int_x^yA\right)T_i\gamma^5
\gamma^0\gamma^{\mu}(\delta(x-y)+\Omega(y,x)).
\lb{4.39} \ee
Like in the abelian case (see (\ref{3.68})) one finds from (\ref{4.37}) that
$<Q>=0$, where $Q$ is the total charge (the first term in (\ref{4.37}) vanishes
after the subtraction of the "free" expectation value, the second term vanishes
since $\Omega$ is traceless in spinor space - see (\ref{4.32})).

In the following we explicitly evaluate the vacuum expectation value of the
chiral charge in two spacetime dimensions. From (\ref{4.39}) we have
\be \langle\rho^5_i(x,y)\rangle=-\frac{1}{2}\mbox{Tr}
 \exp\left(i\int_x^yA\right)T_i\gamma^0\gamma^1(\Omega(y,x)-\Omega_{(0)}(y,x)).
 \lb{4.40} \ee
The trace in (\ref{4.40}) consists actually of two traces: a trace
$\mbox{Tr}_S$ in spinor space and a trace $\mbox{Tr}_C$ in the representation
space of the Lie group. We evaluate the spinor trace by making use of
(\ref{3.37}) and (\ref{4.32}):
\[ -\frac{1}{2}\mbox{Tr}_S\gamma^0\gamma^1\Omega(y,x) =
\frac{im}{\pi}g(y)\sum_n e^{-2\pi i\alpha
n}D^{-n}K_1(m(x-y)+mnL)g^{\dagger}(x). \]
Eq. (\ref{4.40}) then becomes
\bea  \langle\rho^5_i(x,y)\rangle &=& \frac{im}{\pi}\sum_n e^{-2\pi i\alpha n}
   \left(\mbox{Tr}_C
\exp\left(i\int_x^yA\right)T_ig(y)D^{-n}g^{\dagger}(x)\right.
    \nonumber\\
 & & \ \left. -\mbox{Tr}_CT_i\right)K_1(m(x-y)+mnL). \lb{4.41} \eea
The singular terms which arise for $n=0$ cancel. The remaining terms are non
singular in the coincidence limit $x\to y$, and one finds for the expectation
value of the total chiral charge
\bea \langle Q^5_i \rangle &=& \frac{im}{\pi}\int_0^L
    \sum_{n\neq0} e^{-2\pi i\alpha n}
   \left(\mbox{Tr}_C g^{\dagger}(x)T_ig(x)D^{-n}\right.\nonumber\\
   & & \ \left. -\mbox{Tr}_CT_i\right)K_1(mnL). \lb{4.42} \eea
This is the non-abelian version of our earlier result (\ref{3.72}). In the
limit of vanishing mass one finds, using (\ref{4.31}) and $K_1(x)\sim 1/x$,
\bea \langle Q^5_i \rangle &\sim & \frac{2m}{L}\int_0^Ldx
      \mbox{Tr}_C
g^{\dagger}(x)T_ig(x)\sum_ae_ae_a^{\dagger}([\phi_a]+\frac{1}{2}-\phi_a)
\nonumber\\
   & &\ -\mbox{Tr}_CT_i([\alpha]+\frac{1}{2}-\alpha). \lb{4.43} \eea
Note that for semisimple groups the trace of the $T_i$ vanishes.
We emphasize that the currents in the nonabelian theory are
{\em not} gauge invariant quantities but instead transform under the adjoint
representation of the gauge group.

We finally come to the calculation of the vacuum expectation value for the
energy. This closely parallels the discussion of the abelian case which was
discussed in section~3 so that we can be brief in the present case.
The point splitted version of the expectation value now reads, in analogy to
(\ref{3.33}),
\be \langle\Psi\vert H_{\psi}(x,y)\vert\Psi\rangle
   =\frac{1}{2}\mbox{Tr}\int dx\exp\left(-i\int_x^yA\right)h_x(\delta(x-y)
    +\Omega(x,y). \lb{4.44} \ee
We recall that the exponential stands for a path ordered product.
Inspection of the explicit form of the covariance, Eq. (\ref{4.32}), exhibits
that, as in the abelian case, the factors $g(x)$ and $g^{\dagger}(y)$ are
exactly cancelled by the exponential in (\ref{4.44}).  In analogy to
(\ref{3.77}) we then find, after the subtraction of the expectation value for
vanishing external field,
\be \langle H_{\psi}\rangle = \frac{2m}{\pi}\sum_a\sum_{n>0}
        \frac{1}{n}K_1(nmL)\cos(2\pi n\alpha+n\mu_a). \lb{4.45} \ee
In the limit of vanishing mass this becomes
\be  \langle H_{\psi}\rangle_{m=0}=\frac{2\pi}{L}
      \sum_a\left(\alpha+\frac{\mu_a}{2\pi}-[ \alpha+\frac{\mu_a}{2\pi}]
     -\frac{1}{2}\right)^2- \frac{\pi}{6L}N, \lb{4.46} \ee
where $N$ is the number of flavors.

  \chapter {Particle Creation}

\section{Constant electric field in four dimensions}

In this section we demonstrate how the well known expression for the creation
of fermions in a constant external electric field \cite{Sch} can be recovered
in the functional Schr\"odinger picture. The physical picture is the following:
We start with a fermionic vacuum state in the far past (``in - region") and let
it evolve under the influence of the external field, using the Schr\"odinger
equation, into the far future (``out - region"). There we calculate the overlap
with the vacuum in the out - region and interpret the deviation from one as the
probability for particle creation. The state remains, of course, Gaussian but
its exact form (and thus the notion of the vaccum) changes under the evolution
of the external field. It would be physically reasonable to switch on the field
somewhere in the past and switch it off again in the future since no fields
last infinitely long. In the present case of a constant electric field
it will prove advantageous to treat an idealized situation by making use
of the notion of an {\em adiabatic} vacuum state which is approached
in the asymptotic regions. This is possible since $\dot{h}/h$, where
$\dot{h}$ is the time-derivative of the first-quantized Hamiltonian
$h$ (2.5), approaches zero in both the asymptotic past and future.
The concept of adiabatic states is also successfully applied in
traditional discussions of particle creation \cite{Pad} and finds
in particular a fruitful application in quantum theory on curved
spacetimes \cite{BD}.

We thus have for the in - vaccum state
\be \Psi_{in}=N\exp\left(u^{\dagger}\Omega^{in}_{(ad)}u\right), \lb{5.1} \ee
and for the out - vacuum state
\be \Psi_{out}=N\exp\left(u^{\dagger}\Omega^{out}_{(ad)}u\right). \lb{5.2} \ee
The ``adiabatic" covariance $\Omega_{(ad)}$ can be obtained from the ``free"
covariance $\Omega_{(0)}$ (see (2.42)) by replacing the momentum $p$ with
$p+eA$. It turns out to be convenient, in spite of the nonvanishing mass,
to use the chiral representation for the Dirac matrices. The reason is that the
mass terms in the expressions for the covariance become unimportant in the
asymptotic regions. We thus have, instead of (2.42),
\be \Omega_{(ad)}=\frac{1}{\sqrt{\tilde{p}^2+m^2}}\left( \begin{array}{cc}
      -\sigma\cdot\tilde{p} & m \\ m & \sigma\cdot\tilde{p} \end{array}
       \right), \lb{5.3} \ee
where
\be \tilde{p}\equiv (p_x,p_y,p_z+eA_z), \lb{5.4} \ee
and the electric field points in $z$ - direction, ${\bf E}=E{\bf e_z}$, so that
$A_z=Et$. For simplicity we denote the transversal momentum by $p_{\bot}$ so
that $p_{\bot}^2=p_x^2+p_y^2$. It will also be convenient to introduce the
dimensionless quantity
\be \tau\equiv\sqrt{eE}\left(t+\frac{p_z}{eE}\right). \lb{5.5} \ee
We now give the explicit expression for $\Omega_{(ad)}$ in both the asymptotic
past and future. In the limit $\tau\to-\infty$, (5.3) reads ($\sigma_i$ are the
Pauli matrices)
\bea \Omega_{(ad)} &=& \frac{1}{\sqrt{p_{\bot}^2+eE\tau^2+m^2}}
    \left( \begin{array}{cc}-\sigma_{\bot}\cdot
p_{\bot}-\sigma_z\cdot\sqrt{eE}\tau
      & m \\ m & \sigma_{\bot}\cdot p_{\bot}+\sigma_z\cdot\sqrt{eE}\tau
     \end{array}\right)\nonumber\\
 & & \stackrel{\tau\to-\infty}{\longrightarrow}\left(\begin{array}{cc}
       \sigma_z & 0 \\ 0 & -\sigma_z\end{array}\right)
\equiv\Omega_{(ad)}^{in}.
      \lb{5.6} \eea
Analogously,
\be \Omega_{(ad)}^{out}=\left(\begin{array}{cc}-\sigma_z & 0 \\
       0 & \sigma_z \end{array} \right)
       =-\Omega^{in}_{(ad)}. \lb{5.7} \ee
Before we proceed to calculate the pair creation rate according to
the general formula (2.72), we have to discuss one subtlety which arises
through the use of asymptotic vacuum states. As can be immediately seen
by comparing (5.6) and (5.7), the adiabatic covariances
 $\Omega^{out}_{(ad)}$ and $\Omega^{in}_{(ad)}$ differ in their sign.
 Consequently, from the general expression (2.32), the positive
 (negative) frequency eigenfunctions in the far future are the
 negative (positive) frequency eigenfunctions of the far past.  An observer
 in the far future would replace the expansion (2.62) by
 \be \chi_n(t) =\alpha_{nm}^f \chi_m^f+\beta_{nm}^f\psi_m^f
     =\alpha_{nm}^f\psi_m +\beta_{nm}^f\chi_m, \lb{5.8} \ee
 where the superscript $f$ refers to ``far future." Comparing (5.8)
 with (2.62) we see that $\alpha_{nm}^f=\beta_{nm}$ and $\beta_{nm}^f
 =\alpha_{nm}$. Nevertheless, one can still use the expression
 (2.72) to calculate the transition element. The reason is that one
 now has to use $\Omega^{out}_{(ad)} =-\Omega^{in}_{(ad)}$
 instead of $\Omega_0=\Omega^{in}_{(ad)}$ in (2.59). This would
 amount to replace $\beta_{nm}$ in (2.72) by $\alpha_{nm}
 =\beta_{nm}^f$. Thus, the particle creation rate is still given by
 (2.72) with $\beta_{nm}$ replaced by $\beta_{nm}^f$ as it was
 introduced in (5.8) (in the following we will for simplicity omit
 the superscript $f$).

The general expression (2.60) for the covariance $\Omega(t)$
contains, via (2.61), the functions $\chi_n(t)$ which obey
\be i\dot{\chi}_n(t) =h\chi_n(t), \lb{5.9} \ee
where the first-quantized Hamiltonian $h$ is given explicitly by
\be h=\left( \begin{array}{cc}
      \sigma\cdot\tilde{p} & -m \\ -m & -\sigma\cdot\tilde{p} \end{array}
       \right). \lb{5.10} \ee
Note that $h^2 =(p_{\bot}^2 +m^2 +E\tau^2){\bf I}$, and $n$ has to be
replaced by ${\bf p}$. Differentiating (5.9) by $t$ and using (5.9)
again, one arrives at a second order equation for the $\chi_n$.
The first and fourth component of the $\chi_{\bf p}$ obeys
(we omit the index ${\bf p}$ in the following)
\be \left(\frac{d^2}{d\tau^2}+\tau^2+\Lambda+i\right)
\chi_{1,4}=0, \lb{5.11} \ee
while the second and third component obeys
\be \left(\frac{d^2}{d\tau^2}+\tau^2+\Lambda-i\right)
\chi_{2,3}=0. \lb{5.12} \ee
We have introduced in these expressions the quantity
\be \Lambda=\frac{p_{\bot}^2+m^2}{\vert eE\vert}. \lb{5.13} \ee
The discussion is greatly simplified if we treat the
case of two spacetime dimensions first and recover the
four-dimensional case by some simple manipulations from the
final result. Instead of (5.11) and (5.12) we have then
to deal with the equations
\bea & & \left(\frac{d^2}{d\tau^2}+\tau^2+\xi+i\right)
\chi_{1}=0, \lb{5.14} \\
& & \left(\frac{d^2}{d\tau^2}+\tau^2+\xi-i\right)
\chi_{2}=0, \lb{5.15} \eea
where, obviously,
\be \xi=\frac{m^2}{\vert eE\vert}. \lb{5.16} \ee
Since $\chi$ obeys the first-order equation (5.9), the equations
(5.14) and (5.15) cannot be solved independently. If we choose, say,
for $\chi_1$ the general solution of (5.14), we find from (5.9) that
\be \chi_2 =\frac{1}{\sqrt{\xi}}\left(i\frac{d\chi_1}{d\tau}
    -\tau\chi_1\right). \lb{5.17} \ee
The general solution of (5.14) is then given by a sum of parabolic
cylinder functions \cite{GR1}
\be \chi_1 = A_{1}D_{-i\xi/2}[(1+i)\tau]+B_{1}D_{-i\xi/2}
      [-(1+i)\tau]. \lb{5.18} \ee
We now have to impose the boundary condition that $\chi$
approaches a {\em negative frequency eigenfunction} for
$\tau\to-\infty$. For this we need the asymptotic expansion of
(5.18) which reads \cite{GR1}
\bea \chi_1 & \stackrel{\tau\to-\infty}{\sim} &
A_{1}\left(e^{-\frac{i\tau^2}{2}}[(1+i)\tau]^{-\frac{i\xi}{2}}
      \right. \nonumber\\
   & & \left.
-\frac{\sqrt{2\pi}}{\Gamma(\frac{i\xi}{2})}
e^{-\frac{\pi\xi}{2}+\frac{i\tau^2}{2}}
    [(1+i)\tau]^{\frac{i\xi}{2}-1}\right) \nonumber\\
 & & \ +B_{1}e^{-\frac{i\tau^2}{2}}[-(1+i)\tau]^{-\frac{i\xi}{2}}.
\lb{5.19} \eea
The usual definition of positive and negative frequencies involves
the phase of the first-quantized eigenfunctions: For a positive
frequency function the phase decreases with increasing time, while
for a positive frequency function it increases \cite{Pad}. The expression
(5.19) thus should only contain terms proportional to $\exp
(-i\tau^2/2)$. We thus have $A_1=0$ and one is left with
\be \chi_1= B_1D_{-i\xi/2}\left[-(1+i)\right]. \lb{5.20} \ee
 From (5.17) one then gets
\be \chi_2= -\frac{B_1\sqrt{\xi}}{2}(1+i)D_{-i\xi/2-1}
    \left[-(1+i)\right]. \lb{5.21} \ee
We want to normalize the solution $\chi=(\chi_1,\chi_2)^T$.
Since the norm is conserved ($h$ in (5.9) is hermitean), it is
sufficient to perform the normalization in the asymptotic past where
\bea \chi_1 & \stackrel{\tau\to-\infty}{\longrightarrow} &
      B_1 e^{-\frac{i\tau^2}{2}}\vert\tau\vert^{-\frac{i\xi}{2}}
       2^{-\frac{i\xi}{4}}e^{\frac{\pi\xi}{8}}, \lb{5.22} \\
        \chi_2 & \stackrel{\tau\to-\infty}{\longrightarrow} &
        0. \lb{5.23} \eea
Thus, the choice
\be B_1=e^{-\frac{\pi\xi}{8}} \lb{5.24} \ee
yields $\chi^{\da}\chi\equiv \vert\chi_1\vert^2 +\vert\chi_2\vert^2
=1$.

To make use of (5.8) we have to find the positive and negative
frequency functions in the asymptotic future, i.e. for $\tau\to\infty$.
The correctly normalized negative frequency solution $\chi_f$ to
(5.14) and (5.17) reads
\bea \chi_1^f &=& \sqrt{\frac{\xi}{2}}e^{-\frac{\pi\xi}{8}}
     D_{i\xi/2-1}\left[(1-i)\tau\right], \lb{5.25} \\
     \chi_2^f &=& -\frac{i+1}{\sqrt{2}}e^{-\frac{\pi\xi}{8}}
     D_{i\xi/2}\left[(1-i)\right]. \lb{5.26} \eea
 This is easily seen from the asymptotic expansion of the parabolic
 cylinder functions \cite{GR1}. Similarly, the positive frequency
 functions are found to read
 \bea \psi_1^f &=& e^{-\frac{\pi\xi}{8}} D_{-i\xi/2}
      \left[(1+i)\tau\right], \lb{5.27} \\
      \psi_2^f &=& \frac{\sqrt{\xi}}{2}(i+1)e^{-\frac{\pi\xi}{8}}
      D_{-i\xi/2-1}\left[(1+i)\tau\right]. \lb{5.28} \eea
 Making now use of the identity \cite{GR1}
 \be D_{\lambda}(z)= e^{\lambda\pi i}D_{\lambda}(-z)
     +\frac{\sqrt{2\pi}}{\Gamma(-\lambda)}
     e^{\pi(\lambda+1)i/2}D_{-\lambda-1}(-iz),  \lb{5.29} \ee
 we can expand the solution (5.20), (5.21), (5.24) according to (5.8)
 into the asymptotic positive and negative frequency solutions,
 respectively:
 \be \chi(\tau) =\frac{\sqrt{\pi\xi}}{\Gamma(\frac{i\xi}{2}+1)}
     e^{-\frac{\pi\xi}{4}}\chi^f +e^{-\frac{\pi\xi}{2}}\psi^f.
     \lb{5.30} \ee
 The Bogolubov coefficients can be easily read off from this equation,
 \be \alpha=\frac{\sqrt{\pi\xi}}{\Gamma(\frac{i\xi}{2}+1)}
      e^{-\frac{\pi\xi}{4}},\; \beta=e^{-\frac{\pi\xi}{2}}, \lb{5.31} \ee
and it is easily checked that $\vert\alpha\vert^2+ \vert\beta\vert^2=1$.
Finally, one then finds for the matrix element (2.72)
\bea \vert\langle\Psi_1\vert\Psi_2\rangle\vert^2 &=&
     \mbox{det}(1-\vert\beta\vert^2) \nonumber\\
     &=& \exp\mbox{Tr}\ln(1-e^{-\pi\xi}) \nonumber\\
     &=& \exp\left(-\mbox{Tr}\sum_n\frac{1}{n}e^{-\pi n\xi}\right).
     \lb{5.32} \eea
In two dimensions the trace reads
\[ \mbox{Tr}\; \longrightarrow \; \frac{L}{2\pi}
   \int_{eEt_{in}}^{eEt_{out}}dp = \frac{eELT}{2\pi}, \]
where $T\equiv t_{out}-t_{in}$ is the time difference between two
asymptotic times $t_{out}$ and $t_{in}$. This, as well as the
length $L$, has been introduced as an infrared regulator
\cite{Pad}, \cite{Ki}. Thus,
\be \vert\langle\Psi_1\vert\Psi_2\rangle\vert^2 =
    \exp\left(-\frac{eELT}{2\pi}\sum_{n=1}^{\infty}
    \frac{1}{n}e^{-\frac{n\pi m^2}{eE}}\right). \lb{5.33} \ee
(If $eE$ is negative, one has to take its absolute value.)
To find the corresponding expression in four spacetime dimensions,
we have to replace $\xi$ by $\Lambda$, see (5.13). One thus has
\be \vert\beta\vert^2 =e^{-\pi\Lambda}=
    e^{-\frac{\pi(m^2+p_{\bot}^2)}{eE}} \lb{5.34} \ee
and
\[ \mbox{Tr}\; \longrightarrow \; \frac{V}{(2\pi)^3}
    \int_{eEt_{in}}^{eEt_{out}}dp_z \int 2\pi p_{\bot}dp_{\bot}.
    \]
Moreover, one gets an additional factor of $2$ from the discrete part of the
determinant in (5.32) over the spinor indices since one now deals
with four spinors instead of two spinors. Thus,
\bea \vert\langle\Psi_1\vert\Psi_2\rangle\vert^2 &=&
     \exp\left(-2\mbox{Tr}\sum_{n=1}^{\infty}\frac{1}{n}
     e^{-\pi n\Lambda}\right) \nonumber\\
     &=& \exp\left(-\frac{2(eE)^2VT}{(2\pi)^3}\sum_{n=1}^{\infty}
     \frac{1}{n^2}e^{-\frac{n\pi m^2}{eE}}\right). \lb{5.35} \eea
This is in agreement with the classical result of Schwinger \cite{Sch}.

\section{Arbitrary external fields for massless QED in two dimensions}

We now proceed to calculate the vacuum - to - vacuum transition rate (2.58) in
the case of massless fermions for arbitrary external electromagnetic fields in
two spacetime dimensions. In contrast to the previous section we shall assume
that the electric field is switched off for some time $t<t_1$ in the past and
$t>t_2$ in the future. While one can consistently assume that the vector
potential vanishes for $t<t_1$, this is {\em not} possible for $t>t_2$ since
the flux
\be \int_0^L dx\int_{t_1}^{t_2}dtE=\int dxdt\dot{A}=
\int dx\left(A(x,t_2)-A(x,t_1)\right)=
2\pi\varphi(t_2) \lb{5.48} \ee
need not vanish. In fact, this will give rise to the nontrivial features which
will be discussed in this section. We can, however, assume that $A$ does not
depend on $x$ for $t>t_2$.

To determine the covariances $\Omega_1$ and $\Omega_2$
in (2.58) we need to solve the
time-dependent Dirac equation,
\be i\dot{\psi}=h\psi=-i\gamma_5(\partial_x+iA)\psi. \lb{5.49} \ee
We make the ansatz
\be \psi(x,t)=\exp(i\lambda(x,t)+i\delta(x,t)\gamma_5)\psi_0(x,t)  \lb{5.50}
\ee
and choose $\lambda$ and $\delta$ such that $\psi_0$ obeys the free Dirac
equation (without $A$- field). Inserting (\ref{5.50}) into (\ref{5.49}) one
recognizes that this can be achieved if
\bea \dot{\lambda}+\delta'&=& 0,  \nonumber\\
       \lambda'+\dot{\delta}&=& -A. \lb{5.51} \eea
The formal solution reads
\bea \lambda &=& \frac{1}{\Box}A', \nonumber\\
      \delta &=& -\frac{1}{\Box}E. \lb{5.52} \eea
The solution of the free equation for $\psi_0$,
\be i\dot{\psi_0}=-i\gamma_5\partial_x\psi_0, \lb{5.53} \ee
can of course be immediately written down by making use of (3.4) - (3.7)
(we choose $\beta=0$ for simplicity):
\be \psi_{0,n}=\left(\begin{array}{c}\varphi_{0,n}\\
    \chi_{0,n}\end{array}\right) \lb{5.54} \ee
with
\bea \varphi_{0,n} &=& \frac{1}{\sqrt{L}}\exp\left(-ik_n(x+t)\right) \lb{5.55}
\\
        \chi_{0,n} &=& \frac{1}{\sqrt{L}}\exp\left(-ik_n(x-t)\right),
\lb{5.56} \eea
where
\be k_n\equiv\frac{2\pi}{L}(n-\alpha). \lb{5.57} \ee
The positive energy (negative energy) solutions are obtained for $k_n>0$
($k_n<0$) in (5.43) and for $k_n<0$ ($k_n>0$) in (5.44) (recall (3.6) and
(3.7)). The solutions of (5.37) thus read
\be \psi_n(x,t)=\exp(i\lambda+i\delta\gamma_5)\psi_{0,n}. \lb{5.58} \ee
The components of the covariance are calculated in full analogy to Eq. (3.11).
One finds
\be \Omega_+(x,y,t)=\exp(i\lambda(x,t)-i\delta(x,t))\Omega_+^{(0)}(x,y)
      \exp(-i\lambda(y,t)+i\delta(y,t)) \lb{5.59} \ee
and
 \be \Omega_-(x,y,t)=\exp(i\lambda(x,t)+i\delta(x,t))\Omega_-^{(0)}(x,y)
      \exp(-i\lambda(y,t)-i\delta(y,t)), \lb{5.60} \ee
where $\Omega_+^{(0)}$ and $\Omega_-^{(0)}$ are obtained from (3.12) and (3.13)
by setting the $A$- field equal to zero:
\be \Omega_+^{(0)}(x,y)=-\Omega_-^{(0)}(x,y)=
     \frac{i}{L}\exp\left(\frac{2\pi
i}{L}(\alpha-[\alpha]-\frac{1}{2})(x-y)\right)
     \frac{1}{\sin\frac{\pi}{L}(x-y)}. \lb{5.61} \ee
Since $A=0$ for $t<t_1$ one can choose $\lambda=\delta=0$ for $t<t_1$. This
corresponds to the choice of the retarded Green function in (5.40). We thus
have $\Omega=\Omega^{(0)}$ for $t<t_1$.

We now proceed to calculate the overlap (2.58) between the out - vacuum and the
out - state which results from evolving the in - vacuum (which is the free
state) with the Schr\"odinger equation. In the out - region ($t\to\infty$) we
can choose $A$ to be constant. From (5.39) we can choose $\lambda=0$ and
$\delta=-At$. The one particle wave functions (5.46) then read
\be \psi_n(x,t)=\exp(-iAt\gamma_5)\psi_{0,n}(x,t). \lb{5.62} \ee
The out - vacuum is calculated from the wave functions (3.6) and (3.7) for
$A=constant$. As can be recognized from these expressions, $A$ drops out and
one is left with the {\em free} wave functions $\psi_{0,n}$. Does this also
mean that the out - vacuum state is identical with the free vaccum state? This
is {\em not} the case since in the general expression for the covariance, Eq.
(2.32), one has to distinguish between positive and negative energy solutions.
For nonvanishing (even constant) $A$- field this distinction is field-dependent
since the energy values are given by
\be E_n=\pm\frac{2\pi}{L}(n-\phi), \lb{5.63} \ee
where the upper sign is for the right- handed part and the lower sign for the
left- handed part (compare (3.6) and (3.7)). Let us focus in the following on
the right-hand part.
In the expression (2.58) for the overlap we choose for $\Omega_1$ the
covariance which corresponds to the out - vacuum, i. e.,
\be \Omega_1(x,y)=\sum_{n\leq\phi}\psi_{0,n}(x)\psi_{0,n}^{\dagger}(y)-
    \sum_{n>\phi}\psi_{0,n}(x)\psi_{0,n}^{\dagger}(y), \lb{5.64} \ee
where we have included the zero energy eigenfunction in the first sum.
Since $t$ has dropped out in this expression, we have skipped it in the
aruments for the wave functions.
 Since
the phase factor in (5.50) is space-independent, the time-evolved in -
covariance (which plays the role of $\Omega_2$) is just given by
\be \Omega_2(x,y)=\sum_{n\leq\alpha}\psi_{0,n}(x)\psi_{0,n}^{\dagger}(y)-
    \sum_{n>\alpha}\psi_{0,n}(x)\psi_{0,n}^{\dagger}(y). \lb{5.65} \ee
It is clear that this satisfies the time-dependent Schr\"odinger
equation (2.51) trivially with the correct boundary condition
at $t<t_1$.
We then find for the operator product $\Omega_1\Omega_2$ in (2.58)
\bea & & \Omega_1\Omega_2 = \int
dz\left(\sum_{n\leq\phi}\psi_{0,n}(x)\psi_{0,n}^{\dagger}(z)
\sum_{l\leq\alpha}\psi_{0,l}(z)
\psi_{0,l}^{\dagger}(y)\right. \nonumber\\
& & \
+\sum_{n>\phi}\psi_{0,n}(x)\psi_{0,n}^{\dagger}(z)
\sum_{l>\alpha}\psi_{0,l}(z)
\psi_{0,l}^{\dagger}(y)-\sum_{n>\phi}\psi_{0,n}(x)\psi_{0,n}^{\dagger}(z)
 \sum_{l\leq\alpha}\psi_{0,l}(z)\psi_{0,l}^{\dagger}(y) \nonumber\\
 & & \ \  \left.
-\sum_{n\leq\phi}\psi_{0,n}(x)\psi_{0,n}^{\dagger}(z)\sum_{l>\alpha}
   \psi_{0,l}(z)\psi_{0,l}^{\dagger}(y)\right). \lb{5.66} \eea
We may assume without loss of generality that $\phi>\alpha$. The first and
second term in (5.54) give together
\[ \left(\sum_{n\leq\alpha}+\sum_{n>\phi}\right)
    \psi_{0,n}(x)\psi_{0,n}^{\dagger}(y)=\delta(x-y)-\sum_{\alpha<n\leq\phi}
   \psi_{0,n}(x)\psi_{0,n}^{\dagger}(y). \]
The third term vanishes for $\phi>\alpha$, and the last term gives
\[ -\sum_{\alpha<n\leq\phi}
   \psi_{0,n}(x)\psi_{0,n}^{\dagger}(y).  \]
We thus have
\[ \Omega_1\Omega_2=\delta(x-y)-2\sum_{\alpha<n\leq\phi}
   \psi_{0,n}(x)\psi_{0,n}^{\dagger}(y). \]
The determinant in the overlap (2.58) thus contains the operator
\[ {\cal A}\equiv
\frac{1}{2}(1+\Omega_1\Omega_2)=\delta(x-y)-\sum_{\alpha<n\leq\phi}
   \psi_{0,n}(x)\psi_{0,n}^{\dagger}(y). \]
By acting with $\cal A$ on $\psi_{0,k}$ one recognizes that $\cal A$ has a zero
eigenvalue if $\alpha<n\leq\phi$. In this case, therefore, the overlap in
(2.58) {\em vanishes}! This means that the probability for the vacuum to remain
a vacuum is zero -- particles are always created. Since both states $\Psi_1$
and $\Psi_2$ are, however, Gaussians it follows that these states belong to
different Hilbert spaces -- in the case of infinitely many degrees of freedom
the overlap between Gaussians can vanish \cite{Ja}. How can one cope with this
situation? The key to a proper treatment is provided by the observation that
the energy eigenvalues $E_n$ of the first-quantized eigenfunctions exhibit a
{\em spectral flow} -- some of them pass through zero between the in- and out -
region. This is peculiar to the massless case since the energy values $E_n$ do
not change sign for $m\neq0$, see (3.44). As a consequence of the spectral flow
the time - evolved in - state contains, in the out - region, either occupied
positive
energy states or empty negative energy states (for definiteness we assume that
there exist occupied
positive energy states). Our original filling prescription says, however, that
for the vacuum state all positive energy states are empty. To have all states
in the {\em same} Hilbert space (Fock space), one has thus to define the out -
vacuum state by applying as many annihilation operators on the out - Gaussian
as there are occupied energy states, i.e.,
\be \vert 0,out\rangle \equiv
N\prod_{k=1}^{[\varphi]}a_k\exp(u^{\dagger}\Omega_1u).
    \lb{5.67} \ee
Again, $\varphi=(\int_0^LA)/(2\pi)$ is the flux. The time - evolved in - state
can thus be written as
\be \Psi_{in}\stackrel{t\to\infty}{\longrightarrow}N\exp(u^{\dagger}\Omega_1u)
   =\prod_{k=1}^{[\varphi]}a_k^{\dagger}\vert 0,out\rangle. \lb{5.68} \ee
This state thus contains $[\varphi]$ particles with respect to the out -
vacuum,  a result which is of course well known (see, e. g., \cite{Ch}). The
particle creation rate expressed by (5.56) is directly related to the anomaly
in the axial current, and there is a general relationship between the spectral
flow of the first - quantized Dirac hamiltonian, the topological charge, and
the anomalous particle production.
This is very clearly discussed, for example, in \cite{Ja2}.
 The important difference to the previous
subsection is the fact that in the present case a {\em definite} number of
particles has been produced (as given by the flux of the external field),
whereas in the previous case there is a nonvanishing probability for the
production of any number of particles. The Schr\"odinger picture thus provides
us with an intuitive explanation for the anomaly: The filling prescription,
which is crucial for the specification of the ground state, changes in
dependence on the external field. Consequently, the notions of vacuum and
excited states change under the influence
 of the external field.

\chapter{Discussion and Outlook}

The use of different formal approaches to the same theory may not only be
important for making different applications but may also contain the
potentiality to extend the theory into different new directions. In the present
paper we have discussed the functional Schr\"odinger picture for fermionic
fields and some of its applications. Broadly speaking, there are two main
advantages of this. First, the use of wave functionals gives an intuitive
picture of the physics involved, in particular with regard to conceptual
questions. This became especially clear in our discussion of particle creation
and anomalies. Second, this picture may possess technical advantages in some
applications, such as the calculation of expectation values or anomalous
particle production rates. One might therefore expect this picture to be of
some use in other branches of quantum field theory where less results are known
than in QED. In fact, among the next applications we have in mind are fermions
in a gravitational background as well as
 coupled to a quantized gravitational field, especially in the framework of the
new variables in canonical general relativity \cite{As}.
This could shed some light on the final stages of black hole evaporation.
Further possible applications include non-abelian fields in four dimensions
\cite{Lu}, decoherence \cite{Ki2}, the semiclassical approximation \cite{Ki},
bosonization,
as well as the extension to problems where non-Gaussian states play a role.

In the bulk of this paper we have restricted ourselves to the case where the
external electromagnetic field can be treated semiclassically. This is formally
expressed by neglecting terms containing $\delta/\delta {\bf A}(x)$ in the full
Hamiltonian (2.4). We want to relax this restriction now and
 conclude our paper with a brief discussion
 of some
subtleties which arise when the Gauss constraint (2.6) is realized on wave
functionals $\Psi[A,u,u^{\dagger}]$ in the full theory.
 Applying the Gauss operator
\be {\cal G}(x)=\nabla{\bf E}-e\psi^{\dagger}\psi \lb{6.1} \ee
on states $\Psi$ we find, using the realization (2.9) - (2.11) for the field
operators,
\bea {\cal G}(x)\Psi &=& \left(\frac{1}{i}\nabla\frac{\delta}{\delta {\bf
A}}-\frac{e}{2}
      [u^{\dagger}u+\frac{\delta^2}{\delta u\delta u^{\dagger}} \right.
\nonumber\\
     & & \  \left. +u^{\dagger}\frac{\delta}{\delta u^{\dagger}} -
u\frac{\delta}{\delta u}]
       \right)\Psi[A,u,u^{\dagger}]=0. \lb{6.2} \eea
Classically, the Gauss operator generates local gauge transformations. This
also holds in the quantum theory, in the sense that
\be \left[\int dx\lambda(x){\cal G}(x),\psi(y)\right]=e\lambda(y)\psi(y), \
etc.
     \lb{6.3} \ee
with an appropriate test class function $\lambda(x)$. The surprise comes if one
evaluates the expression (6.2) for the Gaussian state (2.14). This yields
\bea {\cal G}(x)\Psi &=& -\frac{1}{2}\int
dydzu_{\alpha}^{\dagger}(y)[\delta(y-x)
    \delta_{\alpha\beta}+\Omega_{\alpha\beta}(y,x)]\times \nonumber\\
    & & \ [\delta(x-z)\delta_{\beta\gamma}-\Omega_{\beta\gamma}(x,z)]
        u_{\gamma}(z)\Psi \neq 0 . \lb{6.4} \eea
Thus, although $\Psi$ is explicitly gauge - invariant, it is {\em not}
annihilated by the Gauss operator. This can also be recognized from a different
perspective. Under an infinitesimal gauge transformation a state $\Psi$ changes
as follows:
\bea \Psi[{\bf A},u,u^{\dagger}] & \to & \Psi[{\bf A},u,u^{\dagger}] - \int
dx\lambda(x)\left(\nabla\frac{\delta}{\delta{\bf A}}+ieu\frac{\delta}{\delta u}
 \right. \nonumber\\
  & & \left. -ieu^{\dagger}\frac{\delta}{\delta u^{\dagger}}\right)\Psi.
\lb{6.5} \eea
The state therefore remains invariant if
\be \left(\frac{1}{i} \nabla\frac{\delta}{\delta{\bf A}}+eu\frac{\delta}{\delta
u}
 -eu^{\dagger}\frac{\delta}{\delta u^{\dagger}}\right)\Psi\equiv\tilde{{\cal
G}}(x)\Psi=0. \lb{6.6} \ee
Obviously, $\tilde{{\cal G}}$ differs from ${\cal G}$. The formal reason is the
fermionic character of the matter fields which allows the realization of the
field operators as in (2.10) and (2.11). In fact, in the bosonic case one has
$\tilde{{\cal G}}\equiv{\cal G}$
\cite{Ki}. Note that the integrated Gauss operator annihilates $\Psi$, i. e.,
\be \int dx{\cal G}(x)\Psi=\int dx\tilde{{\cal G}}(x)\Psi=0. \lb{6.7} \ee
The interpretation of (6.4) was given by Floreanini and Jackiw \cite{FJ}. The
Gauss operator ${\cal G}$ may produce states which lie outside the original
Fock space from which one started, since the space spanned by $u$ and
$u^{\dagger}$ is much bigger than the space obtained from the ground state
through application of the field operators $\psi$ and $\psi^{\dagger}$. They
can only produce polynoms in
\be (1+\Omega)u\equiv u_+, \  \ u^{\dagger}(1-\Omega)\equiv u^{\dagger}_-,
\lb{6.8} \ee
whereas in (6.4) one recognizes their adjoints $u_-$ and $u^{\dagger}_+$:
\be {\cal G}(x)\Psi=-\frac{1}{2}u^{\dagger}_+(x)u_-(x)\Psi. \lb{6.9} \ee
The prescription we impose here is to {\em project} the action of the Gauss
operator back onto the original Fock space,
\[ {\cal G}\to P_F{\cal G}\equiv\frac{1}{4}u_+u^{\dagger}_-{\cal G}. \]
Since the state (6.9) is orthogonal to each state in this space, one has of
course
\be P_F{\cal G}(x)\Psi=0. \lb{6.10} \ee
In particular, one finds that the expectation value of the Gauss operator
vanishes,
$\langle\Psi\vert{\cal G}(x)\Psi\rangle=0$.

There is only one possible obstruction to this prescription: it may happen that
the presence of an anomaly spoils the commutativity of two Gauss operator (this
anomaly should not be confused with the anomaly of the axial current discussed
in the last section). In this case our prescription would lead to a
contradiction since the projected Gauss operators always commute with each
other. An example where such anomalies occur are chiral fermions in an external
electromagnetic field \cite{FJ}. In such a case one {\em cannot} identify a
state $\Psi$ with its projected state, $u_+u^{\dagger}_-\Psi/4$. Here, however,
we deal with Dirac fermions where the anomaly connected with the left - handed
part cancels the corresponding anomaly of the right - handed part. It is thus
perfectly consistent to identify states with their projected version.

In this respect the situation is analogous to the Gupta - Bleuler quantization
of electrodynamics where one can get rid of negative norm states by identifying
states with zero norm.

 We have thus shown that the Gauss operator for fermions can be
consistently interpreted in the functional Schr\"odinger picture if no gauge
violating anomalies are present.

\vspace{0.5cm}

\begin{center}
{\bf Acknowledgement}
\end{center}
We thank the referee for pointing out an error in a preliminary
version of this paper.
This article was supported by the Swiss National Science Foundation.

\end{document}